\documentclass[
  a4paper,
  twocolumn, %appr. journal style! You can also choose 'preprint' or 'twocolumn'
  pra,
  superscriptaddress,
  amsfonts,amsmath,amssymb,
  longbibliography
]{revtex4-2}
\usepackage[
  colorlinks,
  citecolor=blue,
  linkcolor=blue,
  urlcolor=blue
]{hyperref}
\usepackage{graphicx,epstopdf,bm}

\usepackage{braket}
\usepackage[capitalize]{cleveref}
\usepackage{xcolor}
\usepackage[all]{hypcap}

\crefname{section}{Sec.}{Secs.}
\Crefname{section}{Section}{Sections}
\crefname{appendix}{App.}{Appces.}
\Crefname{appendix}{Appendix}{Appendices}

\newcommand{\varfunctional}{J_{\mathrm{cs}}}
\newcommand{\varfunctionalJC}{\tilde{J}_{cs}}

\DeclareMathOperator{\Tr}{Tr}
\newcommand{\e}[1]{\mathrm{e}^{#1}}
\newcommand{\op}[1]{\hat{#1}}
\newcommand{\trace}[2][]{\Tr_{#1}\left[#2\right]}

\newcommand{\cf}{cf.\ }
\newcommand{\eg}{e.g.\ }

\newcommand{\figref}[2][]{Fig.~\hyperref[#2]{\ref*{#2}#1}}
\newcommand{\figureref}[2][]{Figure~\hyperref[#2]{\ref*{#2}#1}}

\newcommand{\spp}{superposition phase}
\newcommand{\opopenone}{\op{\openone}}

\begin{document}

\title{
  Optimizing for an arbitrary Schr\"{o}dinger cat state.
}

%%% Authors %%%%%%%%%%%%%%%%%%%%%%%%%%%%%%%%%%%%%%%%%%%%%%%%%%%%%%%%%%%%%%%%%%%%

\author{Matthias G. Krauss}
\affiliation{
  Theoretische Physik, Universit\"{a}t Kassel,
  Heinrich-Plett-Stra{\ss}e 40, 34132 Kassel, Germany
}
\affiliation{Dahlem Center for Complex Quantum Systems and Fachbereich Physik,
  Freie Universit\"{a}t Berlin, Arnimallee 14, D-14195 Berlin, Germany
}

\author{Christiane P. Koch}
\affiliation{Dahlem Center for Complex Quantum Systems and Fachbereich Physik,
  Freie Universit\"{a}t Berlin, Arnimallee 14, D-14195 Berlin, Germany
}

\author{Daniel M. Reich}
\email{danreich@zedat.fu-berlin.de}
\affiliation{Dahlem Center for Complex Quantum Systems and Fachbereich Physik,
  Freie Universit\"{a}t Berlin, Arnimallee 14, D-14195 Berlin, Germany
}

%%%%%%%%%%%%%%%%%%%%%%%%%%%%%%%%%%%%%%%%%%%%%%%%%%%%%%%%%%%%%%%%%%%%%%%%%%%%%%%%

\begin{abstract}
  We derive a set of functionals for optimization towards an arbitrary cat
  state and demonstrate their application by optimizing the dynamics of
  a Kerr-nonlinear Hamiltonian with two-photon driving. The versatility of our framework
  allows us to adapt our functional towards optimization of maximally entangled cat states,
  applying it to a Jaynes-Cummings model. We identify the strategy of
  the obtained control fields and determine the quantum speed limit as a function
  of the cat state's excitation. Finally, we extend our optimization functionals to
  open quantum system dynamics and apply it to the Jaynes-Cummings model with
  decay on the oscillator. For strong dissipation and large cat radii, we find
  a change in the control strategy compared to the case without dissipation.
  Our results highlight the power of optimal control with functionals
  specifically crafted for complex physical tasks and the versatility of the
  quantum optimal control toolbox for practical applications in the quantum technologies.
\end{abstract}

%%%%%%%%%%%%%%%%%%%%%%%%%%%%%%%%%%%%%%%%%%%%%%%%%%%%%%%%%%%%%%%%%%%%%%%%%%%%%%%%

\maketitle

\section{Introduction}
  \label{sec:introduction}

  Schr\"{o}dinger cat states~\cite{GlauberPR63,Haroche06} constitute an important set of quantum states, which have various
  applications in quantum communication~\cite{vanEnkPRA01}, computation~\cite{CochranePRA99,MirrahimiCRP16,GrimmN20,SuPRA22}, and sensing~\cite{MunroPRA02}.
  In particular, they have been attracting a great deal of attention for the implementation
  of hardware-efficient qubit encodings~\cite{GlancyJOSAB08,MirrahimiCRP16}.
  Recently, these include photonic~\cite{ChenPRA22} and superconducting quantum information~\cite{GrimmN20,SuPRA22}
  architectures due to their intrinsic fault tolerance and suitability for quantum error correction.
  Various approaches have been proposed and implemented for their creation, involving, e.g., a set of fundamental
  logical gates~\cite{VlastakisS13,WangS16}, homodyne detection~\cite{OurjoumtsevS06,OurjoumtsevN07},
  adiabatic protocols~\cite{HatomuraNJP18,ChenPRL21} or quantum reservoir engineering~\cite{ArenzJPBAMOP13,LeghtasS15}.
  Many of these approaches suffer from a long protocol time which in turn limits the speed
  in quantum information applications and makes the cat state generation susceptible to decoherence.

  A powerful tool to obtain fast and robust state preparation protocols is
  quantum optimal control theory~\cite{GlaserEPJD15,KochEQT22}. It aims to
  determine suitable external controls for steering a physical system
  towards a particular goal, often by employing iterative numerical algorithms. The success of such
  algorithms crucially depends on choosing an appropriate optimization functional, i.e., a figure of merit
  encoding the optimization success by a single real number. An obvious choice for state-to-state optimizations
  is the overlap between the state generated by the optimized pulse and the target state. Such functionals
  have also been successfully employed for optimal control of cat state generation~\cite{RojanPRA14,OfekN16,XueNJP22}.
  State overlap functionals only allow for optimization towards a single, specific state whereas
  in many applications it is often sufficient to obtain \textit{any} cat state.
  The actual optimization target is then a set of states and not a single state.
  Since the goal of the optimization is encoded in the functional's extremal
  values~\cite{KochJPCM16}, tailoring the optimization functional to the
  underlying task helps to represent the physical target as faithfully as
  possible.
  This allows for maximal flexibility in finding optimal solutions which might be
  missed by too restrictive functionals.
  Such specifically crafted functionals for complex optimization tasks have
  proven to be very successful, e.g., in the optimization of individual optical cycles
  in laser cooling of molecules~\cite{ReichNJP13} or the optimization towards sets of entangling
  quantum gates~\cite{MullerPRA11,WattsPRA15} instead of specific gates.

  Here we derive an optimization functional which allows for optimization towards the entire set of cat
  states instead of only a specific element. By engineering a set of functional terms, which individually check for
  all desired properties of the target state, we are able to construct a composite functional which takes on its extremal, optimal
  value if and only if a cat state is obtained. To illustrate our functional we show its application in
  creating cat states in a simple Kerr-nonlinear oscillator with two-photon driving.

  We further demonstrate the power of our framework by studying the optimization of
  maximally entangled cats in a bipartite system. Such states are important in quantum sensing
  applications~\cite{MunroPRA02,PenasaPRA16} or to implement so-called flying qubits~\cite{WangSA22}.
  To perform optimizations for such a problem, we refine our functional such that it also checks whether
  maximal entanglement between the two subsystems is generated at final time.
  Optionally, the radius of the cat, i.e., the displacement of the corresponding coherent state, may also be prescribed.
  This option is particularly relevant when the system is subject to decay because the maximally achievable
  radius will be determined by a balance between the coherent mechanism that allows for preparing the cat,
  \eg a Kerr non-linearity, and the decay~\cite{HePRA09}.

  Finally, in the present era of noisy quantum devices, it is imperative to also account for the sources
  of noise when deriving pulse shapes for practical applications.
  Decay and dephasing processes due to couplings between the quantum system and its environment
  constitute a major such source, and much effort has been devoted to adapting optimal control theory
  to open quantum systems~\cite{KochJPCM16}.
  To account for this impact during the optimization and search for control strategies that
  can avoid or mitigate this effect, it is imperative to include the effect of the environment in
  the model and adapt the optimization functional. We do this here by rewriting our optimization functional in the
  density matrix formalism and employ a Markovian master equation to describe the noisy quantum dynamics.

  The paper is organized as follows. We begin in \cref{sec:framework-for-targeting-cat-states} by briefly introducing
  cat states as well as the basic framework of optimal control theory, using
  Krotov's algorithm as an example. At the end of \cref{sec:framework-for-targeting-cat-states} we present
  our construction of the functional for optimization towards arbitrary cat states and arbitrary maximally entangled cat states
  as well as the necessary modifications when employing our functionals for open quantum systems.
  \Cref{sec:optimization-results-for-kerrphoton-driving} illustrates the application of the cat state
  functional for optimization in a Kerr-nonlinear oscillator.
  In \cref{sec:optimization-towards-entangled-cat-states-in-a-bipartite-system}
  we present results for an example optimization towards maximally entangled cat states in a
  Jaynes-Cummings model for both coherent and dissipative dynamics.
  Finally, we also discuss which insight can be drawn from the optimization results regarding the role
  of the excitation of the cat state and the quantum speed limit and compare the performance of
  the optimized pulses we obtain with and without taking the decay into account.
  We conclude in \cref{sec:conclusions}.

%###############################################################################
%###############################################################################

\section{Framework for targeting cat states}
  \label{sec:framework-for-targeting-cat-states}

  Cat states are defined as a superposition of two coherent states $\ket{\alpha}$
  where the complex-valued parameter $\alpha$ describes the displacement in phase space, respectively the excitation,
  of the coherent states with respect to the ground state $\ket{0}$~\cite{GlauberPR63,Haroche06}.
  A general form of a cat state is given as
  \begin{equation}
    \label{eq:arb_cat_state}
    \ket{\psi_\mathrm{cat}}=\frac{1}{\mathcal{N}_\varphi}
      \Big(\ket{\alpha} + \e{i\varphi} \ket{-\alpha}\Big)\,,
  \end{equation}
  with $\mathcal{N}_\varphi=\sqrt{2\big(1
  + \e{-2|\alpha|^2}\cos(\varphi)\big)}$ accounting for normalization.
  In the following, we refer to the relative phase $\varphi$ between the
  two coherent states, as ``\spp{}''.
  Note that it is also possible to consider superpositions of more than two
  coherent states, which are commonly referred to as
  multicomponent cat states~\cite{MilburnPRA86,YurkePRL86a,BergmannPRA16}.
  However, in this paper we use the term ``cat state'' exclusively for two-component
  superpositions as the focus of our study.

  Before we move to derive functionals optimizing towards an arbitrary cat state
  we briefly review Krotov's method~\cite{Krotov95,KonnovARC99,GoerzSP19}
  which we employ for all numerical optimizations performed in this work.

  %%%%%%%%%%%%%%%%%%%%%%%%%%%%%%%%%%%%%%%%%%%%%%%%%%%%%%%%%%%%%%%%%%%%%%%%%%%%%%
  \subsection{Optimization algorithm}
    \label{ssec:optimization-algorithm}
  %%%%%%%%%%%%%%%%%%%%%%%%%%%%%%%%%%%%%%%%%%%%%%%%%%%%%%%%%%%%%%%%%%%%%%%%%%%%%%

  Krotov's method~\cite{Krotov95,KonnovARC99,GoerzSP19} is an iterative, monotonically convergent
  optimization algorithm using gradient information to achieve convergence. The optimization
  target typically consists of two parts,
  \begin{equation}
    \label{eq:abstract-oct-func}
    J = J_T[\psi(T)]
      + \int_0^T J_t\big[\psi(t), \varepsilon(t)\big] \,\mathrm{d}t\,,
  \end{equation}
  where we assume a control problem described by a single Hilbert space
  state $\psi(t)$ and a single control field $\varepsilon(t)$, commonly
  realized by external electromagnetic pulses.
  The first part of $J_T$ depends only on the state at final time $T$ and
  encodes the target to be reached at the end of the control pulse, whereas
  the intermediate-time functional $J_t$ describes further costs.
  In Krotov's algorithm, the following cost functional is usually employed,
  \begin{equation}
    \label{eq:j-oct-func}
    J_t\big[\psi(t), \varepsilon(t)\big] =
      \frac{\lambda_a}{S(t)}\big[
        \varepsilon(t)-\varepsilon_\mathrm{ref}(t)
      \big]^2\,,
  \end{equation}
  with the shape function $S(t)$ ensuring that the optimized field is smoothly
  switched on and off. The reference field $\varepsilon_\mathrm{ref}(t)$ is commonly
  taken to be the field from the previous iteration.
  This choice allows the parameter $\lambda_a$ to tune the step size of
  the optimization algorithm by penalizing large changes in the control field
  between iteration steps~\cite{PalaoPRA03}.

  The update equation for the pulse in the iteration $k+1$ of the algorithm
  is given by~\cite{PalaoPRA03,ReichJCP12,GoerzNJP14,GoerzSP19,BasilewitschPRR20,GoerzNJP21}
  \begin{widetext}
    \begin{align}
      \label{eq:krotov_update_eq}
      \varepsilon^{(k+1)}(t) &= \varepsilon^{(k)}(t) +
          \frac{S(t)}{\lambda_a} \mathfrak{Im}\bigg\{
            \bigg\langle \chi^{(k)}(t)
            \bigg|
            \frac{\partial\op{H}}{\partial\varepsilon}
              \Big|_{\varepsilon^{(k+1)}(t)}
            \bigg|
            \psi^{(k+1)}(t)
            \bigg\rangle
          \bigg\} \qquad \text{and} \\[\baselineskip]
      \label{eq:krotov_update_eq_liouville}
      \varepsilon^{(k+1)}(t) &= \varepsilon^{(k)}(t) +
          \frac{S(t)}{\lambda_a} \mathfrak{Re}\bigg\{
            \Big\langle \op{\chi}^{(k)}(t),\
            \frac{\partial\mathcal{L}}{\partial\varepsilon}
            \Big|_{\varepsilon^{(k+1)}(t)} \op{\rho}^{(k+1)}(t)\Big\rangle
          \bigg\}\,,
    \end{align}
  \end{widetext}
  for coherent and dissipative dynamics, respectively. For the dissipative
  case we use density matrices instead of Hilbert space states to describe
  the state of our system and we assume that the time evolution is generated
  by a Liouvillian superoperator $\mathcal{L}$ - a detailed discussion of this
  framework in the context of Krotov's method can be found in Refs.~\cite{GoerzNJP14,GoerzNJP21}.
  $\frac{\partial\op{H}}{\partial\varepsilon}$ is the derivative of the
  Hamiltonian with respect to the control pulse $\varepsilon$, while
  $\frac{\partial\mathcal{L}}{\partial\varepsilon}$ describes the derivative of
  the Liouvillian superoperator $\mathcal{L}$  with respect to the pulse.

  The $\ket{\chi^{(k)}(t)}$ are often called costates. They are propagated backward in time
  according to the equation of motion,
  \begin{align}
    \frac{\mathrm{d}}{\mathrm{d}t}\ket{\chi^{(k)}(t)}
      &= \op{H}\big[\varepsilon^{(k)}(t)\big]
        \ket{\chi^{(k)}(t)}\,,
    \label{eq:krotov_chi_prop}
  \end{align}
  with the boundary condition
  \begin{align}
    \ket{\chi^{(k)}(T)} &= -\nabla_{\bra{\psi}} J_T
    \Big|_{t=T}\,.
    \label{eq:krotov_chi_bc}
  \end{align}
  The states $\ket{\psi^{(k+1)}(t)}$ are obtained by solving the equation of motion
  \begin{subequations}
    \label{eq:krotov_psi_eom}
    \begin{align}
      \frac{\mathrm{d}}{\mathrm{d}t}\ket{\psi^{(k+1)}(t)}
        &= \op{H}\big[\varepsilon^{(k+1)}(t)\big]\ket{\psi^{(k+1)}(t)}
      \label{eq:krotov_psi_prop}
      \\
      \ket{\psi^{(k+1)}(0)} &= \ket{\psi_0}\,,
      \label{eq:krotov_psi_bc}
    \end{align}
  \end{subequations}
  where $\ket{\psi_0}$ is the initial state of the system.

  Similarly to \cref{eq:krotov_chi_prop,eq:krotov_chi_bc,eq:krotov_psi_eom},
  equations of motion for the dissipative dynamics, governed by $\mathcal{L}$,
  are given as
  \begin{subequations}
    \label{eq:krotov_diss_eom}
    \begin{align}
      \frac{\mathrm{d}}{\mathrm{d}t}\op{\rho}^{(k+1)}(t)
        &= \mathcal{L}\big[\varepsilon^{(k+1)}(t)\big]\op{\rho}^{(k+1)}(t)
      \label{eq:krotov_codm_prop}
      \\
      \frac{\mathrm{d}}{\mathrm{d}t}\op{\chi}^{(k)}(t)
        &= \mathcal{L}^{\dagger}\big[\varepsilon^{(k)}(t)\big]
          \op{\chi}^{(k)}(t),
      \label{eq:krotov_dm_prop}
      \\
      \label{eq:krotov_codm_bc}
      \op{\rho}^{(k+1)}(0) &= \op{\rho}_0,
      \\
      \label{eq:krotov_dm_bc}
      \op{\chi}^{(k)}(T) &= -\nabla_{\op{\rho}} J_T
      \Big|_{\op{\rho}^{(k)}(T)},
    \end{align}
  \end{subequations}
  where $\mathcal{L}\big[\varepsilon^{(k)}(t)\big]$ is the Liouvillian
  with the set of controls $\varepsilon^{(k)}(t)$ for the $k$th iteration and
  the initial state $\op{\rho}_0$.

  Together, these equations define the iterative optimization algorithm which is
  started by picking a guess pulse $\varepsilon^{(0)}(t)$.

  %%%%%%%%%%%%%%%%%%%%%%%%%%%%%%%%%%%%%%%%%%%%%%%%%%%%%%%%%%%%%%%%%%%%%%%%%%%%%%
  \subsection{Functional targeting a cat state}
    \label{ssec:functional-targeting-a-cat-state}
  %%%%%%%%%%%%%%%%%%%%%%%%%%%%%%%%%%%%%%%%%%%%%%%%%%%%%%%%%%%%%%%%%%%%%%%%%%%%%%

  To construct a final time functional which allows for optimizing
  towards the set of cat states described by
  \cref{eq:arb_cat_state}, we use the fact that the variance
  of an operator $\op{O}$ in a state $\ket{\psi}$,
  \begin{equation}
    \label{eq:variance}
    \Delta_{{\psi}}\op{O}
      = \braket{\psi|\op{O}^\dagger\op{O}|\psi}
        - \big|\braket{\psi|\op{O}|\psi}\big|^2,
  \end{equation}
  is zero if and only if $\ket{\psi}$ is an eigenstate of that operator.
  Coherent states are eigenstates of the annihilation operator,
  $\op{a}\ket{\alpha} = \alpha\ket{\alpha}$.
  As a result, all cat states as defined in \cref{eq:arb_cat_state}
  are eigenstates of $\op{a}^2$.
  Due to this property, we use the variance of $\op{a}^2$,
  \begin{equation}
    \label{eq:var_func}
    \varfunctional(\psi) = \Delta_{{\psi}}\op{a}^2
      = \Braket{\psi|\big(\op{a}^\dagger\big)^2\op{a}^2|\psi}
        - \Big|\Braket{\psi|\op{a}^2|\psi}\Big|^2,
  \end{equation}
  as a starting point for the functional.
  However, cat states are not the only eigenstates of $\op{a}^2$, such that
  a vanishing variance is a necessary but not sufficient condition to
  identify an element from the set of cat states. Rather, all states of shape
  \begin{equation}
    \label{eq:general_coh_suppos}
    \ket{\psi_{a^2}} = c_0\ket{\alpha} + c_1\ket{-\alpha}
  \end{equation}
  with $|c_0|^2+|c_1|^2 + 2\,\mathfrak{Re}\{c^\ast_0c_1\}\braket{\alpha|-\alpha}=1$ and
  $c_i \in \mathbb{C}$, lead to $\varfunctional(\psi)=0$.
  To obtain an expression whose minimal value is both necessary and sufficient
  to identify a cat state, we construct a composite functional with multiple terms,
  \begin{equation}
    \label{eq:j_t_original}
    J_\mathrm{T}(\psi) = \varfunctional(\psi) + J_\mathrm{cat}(\psi).
  \end{equation}
  The first term is minimized if and only if the state is a superposition of
  $\ket{\alpha}$ and $\ket{-\alpha}$, \cf \cref{eq:general_coh_suppos}
  and the second term attains its minimum
  if and only if the desired subset of these superposition states is reached.
  We refer to $\varfunctional(\psi)$ as ``coherent state term'' and to
  $J_\mathrm{cat}(\psi)$ as ``cat term'', due to their purpose in the overall
  functional. The specific form of the latter depends on which set of states
  should be targeted.

  For example, optimizing towards the set of even and odd cat states,
  \begin{align}
    \label{eq:even_odd_cat}
    \ket{\psi_\mathrm{cat}^\pm} \propto \ket{\alpha} \pm \ket{-\alpha}\,,
  \end{align}
  respectively, can be achieved by choosing $J_\mathrm{cat}$ as
  \begin{equation}
    \label{eq:j_cat_parity_original}
    J_{\mathrm{cat},\pm}(\psi) =
      1-|\langle\psi|\op{\Pi}_{\pm}|\psi\rangle|^2\,,
  \end{equation}
  where
  \begin{align}
    \label{eq:parity_projector}
    \op{\Pi}_{+} = \sum_{j\text{ even}} \ket{j}\bra{j}\,, &&
    \op{\Pi}_{-} = \sum_{j\text{ odd}} \ket{j}\bra{j}\,.
  \end{align}
  are the projectors onto the eigenspaces of the parity operator.
  $\op{\Pi}_{+}$ and $\op{\Pi}_{-}$ project onto even and odd cat states with
  a \spp{} $\varphi$ of 0(even) and
  $\pi$(odd), respectively, while not imposing any restriction on the value of
  $\alpha$.

  Another choice of $J_\mathrm{cat}$ which allows to also leave the \spp{}
  $\varphi$ free, is
  \begin{equation}
    \label{eq:j_cat_arb_original}
    J_{\mathrm{cat},\varphi}(\psi) =
      \Big(\bar{a}^\ast \braket{\psi|\op{a}|\psi}
            + \bar{a}\braket{\psi|\op{a}^\dag|\psi}\Big)^2\,,
  \end{equation}
  with $\bar{a} \equiv \sqrt{\Braket{\psi|\op{a}^2|\psi}}$.
  We prove in \cref{app:derivation-spp-functional}, that
  $J_{\mathrm{cat},\varphi}(\psi)$ is indeed minimal if and only if $|c_0|^2
  = |c_1|^2$ in \cref{eq:general_coh_suppos}.
  With this, the functional term does not impose any restrictions on $\varphi$
  and allows to optimize towards the general set of cat states defined in
  \cref{eq:arb_cat_state}.
  %-----------------------------------------------------------------------------

  A property shared by the functional terms in \cref{eq:var_func} and
  \cref{eq:j_cat_arb_original} is that their value is strongly suppressed for $|\alpha|
  \rightarrow 0$.
  This is due to the summands in both terms being proportional
  to $|\alpha|^4$ and thus tending towards zero as $|\alpha| \rightarrow 0$.
  Therefore, the functional value is reduced for smaller values of $\alpha$.
  Such a behavior leads to an artificial pull towards small values of
  $|\alpha|$, which can be problematic since many applications of cat states
  rely on a large displacement of the two coherent states in phase space,
  corresponding to larger $|\alpha|$.
  Examples for benefits of large displacements are increased sensing
  accuracy~\cite{CochranePRA99} or an improved robustness against errors
  in quantum information applications~\cite{MirrahimiNJP14}.
  The tendency towards small $|\alpha|$ can be amended by normalizing the
  coherent state term,
  \begin{equation}
    \label{eq:var_a_func_norm}
    \varfunctional(\psi)
      = 1 - \frac{
        \Big|\Braket{\psi|\op{a}^2|\psi}\Big|^2
      }{
        \Braket{\psi|\big(\op{a}^\dagger\big)^2\op{a}^2|\psi}
      }\,,
  \end{equation}
  and the cat term for arbitrary \spp{}s,
  \begin{equation}
    \label{eq:j_cat_arb_original_norm}
    J_{\mathrm{cat},\varphi}(\psi) =
      \frac{\braket{\psi|\op{a}|\psi}}{\bar{a}}
      + \frac{\braket{\psi|\op{a}^\dag|\psi}}{\bar{a}^\ast}\,.
  \end{equation}
  For the optimizations presented below, we always use the normalized expressions defined in
  \cref{eq:j_cat_arb_original_norm,eq:var_a_func_norm}.

  %%%%%%%%%%%%%%%%%%%%%%%%%%%%%%%%%%%%%%%%%%%%%%%%%%%%%%%%%%%%%%%%%%%%%%%%%%%%%%
  \subsection{Functional targeting an entangled cat state}
    \label{ssec:functional-targeting-an-entangled-cat-state}
  %%%%%%%%%%%%%%%%%%%%%%%%%%%%%%%%%%%%%%%%%%%%%%%%%%%%%%%%%%%%%%%%%%%%%%%%%%%%%%

  The cat state functionals can be further
  adapted to more complex optimization targets. In this section
  we show how to extend our framework to target maximally entangled cat states in a
  bipartite system. Specifically, we consider a harmonic oscillator coupled to an atom described
  as a qubit -- an ubiquitous physical setup in the field of
  cavity quantum electrodynamics and circuit quantum electrodynamics~\cite{Haroche06}.
  The corresponding optimization targets are given by states of the form
  \begin{equation}
    \label{eq:ent_cat_state}
    \ket{{\Psi}_{\mathrm{cat}}}
      = \frac{1}{\sqrt{2}}\big(
          \ket{b_+} \otimes \ket{\psi_\mathrm{cat}^+}
          + \ket{b_-} \otimes \ket{\psi_\mathrm{cat}^-}
        \big)\,,
  \end{equation}
  where $\ket{\psi_\mathrm{cat}^\pm}$ is defined in \cref{eq:even_odd_cat} and
  $\ket{b_\pm}$ denotes an arbitrary orthonormal basis of the qubit.
  In the spirit of allowing maximal flexibility, we again aim to derive a functional to
  optimize towards arbitrary maximally entangled cat states, i.e., the entire set of such states.
  In particular, we do not want to impose any restrictions on the basis states of
  the qubit involved in the superposition.
  However, if a specific basis on the qubit is desired, the functional can be easily adapted
  by introducing an additional term to the optimization functional.
  For example, the parity operators can be employed to project out one of the two
  summands constituting the entangled cat state in \cref{eq:ent_cat_state}.
  Then, it is straightforward to fix the state $\ket{b_\pm}$ by
  calculating the overlap with the desired basis state.

  To generalize the final time functional, \cref{eq:j_t_original}, to an entangled
  cat state in a bipartite system, we modify the operator used to calculate the
  variance,
  \begin{equation}
    \label{eq:bipartite_lowering_op}
    \op{a} \rightarrow \op{A}\equiv \opopenone\otimes\op{a}\,.
  \end{equation}
  Thus, the coherent state term is replaced by
  \begin{equation}
    \label{eq:var_func_product_simple}
    \varfunctionalJC(\Psi)
      = \Delta_{{\Psi}}\op{A}^2\,.
  \end{equation}
  Similar to the discussion in \cref{ssec:functional-targeting-a-cat-state}, it
  is insufficient to use only
  the coherent state functional $\varfunctionalJC(\Psi)$ since it takes on
  its minimal values not only for the desired set of states defined in
  \cref{eq:ent_cat_state}, but for a larger set of states given by
  \begin{equation}
    \label{eq:general_state_arb_qubit}
    \ket{\Psi_{\mathrm{ent}}}
      = d_0\big(
          \ket{g} \otimes \ket{\psi_{\mathrm{0,a^2}}}
        \big)
        + d_1\big(
          \ket{e} \otimes \ket{\psi_{\mathrm{1,a^2}}}
        \big)\,,
  \end{equation}
  with $\ket{\psi_{\mathrm{j,a^2}}}$ being eigenstates of $\op{a}^2$
  (\cf~\cref{eq:general_coh_suppos}) and $d_i
  \in \mathbb{C}$ with $|d_0|^2 + |d_1|^2 = 1$.
  To amend this, we once again construct a composite functional, adding an additional term
  to restrict the set of states to exactly those of the form of $\ket{{\Psi}_{\mathrm{cat}}}$.
  We accomplish this by exploiting the fact that the targeted states $\ket{{\Psi}_{\mathrm{cat}}}$ are
  maximally entangled. This means that the reduced states of the harmonic oscillator, respectively the qubit,
  take on the minimal purity value~\cite{Jaeger07} of $\mathcal{P}=0.5$.
  Note that the more straightforwards choice of using the von Neumann entropy as
  measure for entanglement cannot be used as functional.
  Its derivative with respect to $\bra{\psi}$, which is needed for the
  calculation of the costates, \cf \cref{eq:krotov_chi_bc}, exhibits
  singularities, which can easily lead to numerical instabilities.
  For this reason we employ the following cat term tracking the subsystem
  purity,
  \begin{equation}
    \label{eq:j_purity_arb_entangled_cats}
    {J}_{\mathrm{cat}}(\Psi) = 2\Tr(\op{\rho}^2_\mathrm{HO}) - 1
                          = 2\Tr(\op{\rho}^2_\mathrm{qubit}) - 1\,,
  \end{equation}
  where
  $\op{\rho}_\mathrm{HO}=\trace[\mathrm{qubit}]{\op{\rho}}$ and
  $\op{\rho}_\mathrm{qubit}=\trace[\mathrm{HO}]{\op{\rho}}$.
  $\trace[\mathrm{qubit}]{\cdot}$ and $\trace[\mathrm{HO}]{\cdot}$ are the
  partial traces corresponding to the qubit and the harmonic oscillator,
  respectively.
  Note that we have used the fact that $\op{\rho}$ is a pure state in the second
  equality.
  This assumption is valid for coherent dynamics. We introduce an extension of our
  formalism to dissipative dynamics in the next section.
  Furthermore,
  as is commonly done in optimal control, we have renormalized
  ${J}_{\mathrm{cat}}(\Psi)$ such that it takes on values between zero
  and one.
  For the cat term in \cref{eq:j_purity_arb_entangled_cats}, we explicitly
  calculate the costates in \cref{app:calculation-co-state}.

  The cat term in \cref{eq:j_purity_arb_entangled_cats} ensures both that the eigenstates
  of $\op{a}^2$, $\ket{\psi_{\mathrm{j,a^2}}}$ in the entangled cat state superposition are
  orthogonal to each other, i.e.,
  $\braket{\psi_{\mathrm{0,a^2}}|\psi_{\mathrm{1,a^2}}}=0$, and that the prefactors in \cref{eq:general_state_arb_qubit} are the
  same, i.e., $|d_0|=|d_1|$.
  Thus, the combined functional,
  \begin{equation}
    \label{eq:functional_completely_entangled_state}
    {J}_T(\Psi) =
      \varfunctionalJC(\Psi)+{J}_{\mathrm{cat}}(\Psi)\,,
  \end{equation}
  takes on its minimal value only for states
  \begin{equation}
    \label{eq:general_state_arb_qubit_completely_entangled}
    \ket{\Psi_{\mathrm{ent}}}
      = \frac{1}{\sqrt{2}}\big(
          \ket{b_0} \otimes \ket{\psi_{\mathrm{0,a^2}}} +
          \ket{b_1} \otimes \ket{\psi_{\mathrm{1,a^2}}}
        \big)\,,
  \end{equation}
  with $\braket{\psi_{\mathrm{0,a^2}}|\psi_{\mathrm{1,a^2}}} = 0$ and
  $\ket{b_j}$ being arbitrary orthogonal basis states of the qubit.
  Indeed, the states in \cref{eq:general_state_arb_qubit_completely_entangled}
  are equivalent to those in \cref{eq:ent_cat_state}, which we prove in
  \cref{app:proof-ent-cat-equiv}.

  %%%%%%%%%%%%%%%%%%%%%%%%%%%%%%%%%%%%%%%%%%%%%%%%%%%%%%%%%%%%%%%%%%%%%%%%%%%%%%
  \subsection{Dissipation adapted functional}
    \label{ssec:dissipation-adapted-functional}
  %%%%%%%%%%%%%%%%%%%%%%%%%%%%%%%%%%%%%%%%%%%%%%%%%%%%%%%%%%%%%%%%%%%%%%%%%%%%%%

  As in the previous section, optimization will target an entangled cat
  state,
  \begin{equation}
    \label{eq:ent_cat_state_diss}
    \ket{{\Psi}_{\mathrm{cat}}}
      = \frac{1}{\sqrt{2}}\big(
          \ket{b_+} \otimes \ket{\psi_\mathrm{cat}^+}
          + \ket{b_-} \otimes \ket{\psi_\mathrm{cat}^-}
        \big)\,.
  \end{equation}
  When considering open system evolution, we need to adapt the optimization
  functional derived in \cref{ssec:functional-targeting-an-entangled-cat-state}
  to density operators.
  The coherent state term $J_\mathrm{cs}$ is simply defined in terms of the
  variance. For open quantum systems, it becomes
  \begin{equation}
    \label{eq:var_a_func_norm_diss}
    J_\mathrm{cs}(\op{\rho})
      = \Tr{\Big[\big(\op{A}^\dagger\big)^2\op{A}^2\op{\rho}\Big]}
        -
        \Big|\Tr{\big[\op{A}^2\op{\rho}\big]}\Big|^2\,,
  \end{equation}
  where $\op{A}\equiv \opopenone\otimes\op{a}$.
  Using the definition of the partial trace, this is equivalent to
  \begin{equation}
    \label{eq:var_func_product_simple_alt}
    J_\mathrm{cs}(\op{\rho})
      = \trace{\op{a}^2\op{\rho}_{\mathrm{HO}}\big(\op{a}^\dagger\big)^2}
        -
        \Big|\trace{\op{a}^2\op{\rho}_{\mathrm{HO}}}\Big|^2,
  \end{equation}
  where $\op{\rho}_{\mathrm{HO}} = \trace[\mathrm{qubit}]{\op{\rho}}$ is the
  reduced state of the HO.
  Analogously to \cref{ssec:functional-targeting-a-cat-state}, the coherent
  state term can be normalized to counter the tendency towards
  $|\alpha|\rightarrow 0$,
  \begin{equation}
    \label{eq:var_func_product_simple_alt_norm}
    J_\mathrm{cs}(\op{\rho})
      = 1 - \frac{
          \Big|\trace{\op{a}^2\op{\rho}_{\mathrm{HO}}}\Big|^2
        }{
          \trace{\op{a}^2\op{\rho}_{\mathrm{HO}}\big(\op{a}^\dagger\big)^2}
        }\,.
  \end{equation}
  Finally, $J_\mathrm{cat}$ ensures equal weights in the
  superposition~\eqref{eq:ent_cat_state}, which in
  \cref{ssec:functional-targeting-an-entangled-cat-state} is achieved by making
  use of the subspace purity.
  This cannot be so straightforwardly generalized to open system evolution.
  When optimizing coherent dynamics, the purity of either of the
  subsystems suffices to determine if the final state is maximally entangled.
  However, in case of non-unitary evolution, the subsystem purity is not only
  reduced by entangling the two systems, but also decreases due to dissipation.
  This can for example lead to different values for the two subsystem purities.
  We thus express $J_\mathrm{cat}(\op{\rho})$ in terms of
  the mutual information which is symmetric and, in general, defined as
  \begin{equation}
    \label{eq:vn_mutual_info}
    \mathcal{I}\big(\text{HO}\!:\!\text{qubit}\big)
      = S(\op{\rho}_\mathrm{HO}) + S(\op{\rho}_\mathrm{qubit}) - S(\op{\rho})
  \end{equation}
  with $S(\op{\rho})$ the von Neumann entropy, $S(\op{\rho}) = -\Tr[\op{\rho}\ln\op\rho]$.
  We use the mutual information to assess the correlation of the optimized state
  which helps to steer the optimization towards our target state even in the
  early stages of the optimization when purity is low. Moreover, the mutual information
  takes on its maximal value for our maximally entangled target states and for pure
  states it is proportional to the entanglement entropy~\footnote{In bipartite systems
  where the two subsystems possess different dimensionality, it is possible to
  obtain maximal entanglement also for certain
  mixed state~\cite{LiQIC12}. However, the term $J_\mathrm{cs}$ in our optimization
  functional becomes minimal only if the state of the harmonic oscillator is
  restricted to a two-dimensional subspace spanned by $\ket{\pm\alpha}$. Thus,
  our optimization is steered towards the case where both constituent spaces
  are effectively two-dimensional, thereby ensuring that our combined
  functional takes on its minimal values only for pure states}.
  Similar to the coherent case, the von Neumann entropy is inconvenient since
  the derivative of $S(\op{\rho})$, which is needed for calculation of costate
  in \cref{eq:krotov_dm_bc}, is not always defined.
  To avoid this numerical problem, we use the linear
  entropy~\cite{ZurekPRL93,ManfrediPRE00} instead,
  \begin{equation}
    \label{eq:lin_entropy}
    S_\mathrm{lin}(\op{\rho})=1-\mathcal{P}(\op{\rho})
      =1-\Tr[\op{\rho}^2]\,.
  \end{equation}
  Accounting for the fact that all terms in the final time functional will be
  minimized, we obtain
  \begin{align}
    \label{cat_term}
    J_\mathrm{cat}(\op{\rho})
       &= 1 - \Big(S_\mathrm{lin}(\op{\rho}_\mathrm{HO})
          + S_\mathrm{lin}(\op{\rho}_\mathrm{qubit})
          - S_\mathrm{lin}(\op{\rho})\Big)
    \notag\\
       &= \mathcal{P}(\op{\rho}_\mathrm{HO})
          + \mathcal{P}(\op{\rho}_\mathrm{qubit})
          - \mathcal{P}(\op{\rho})\,,
  \end{align}
  replacing \cref{eq:j_purity_arb_entangled_cats} in the presence
  of dissipation.

%###############################################################################
%###############################################################################

\section{Optimization Results for a Kerr-nonlinear resonator}
  \label{sec:optimization-results-for-kerrphoton-driving}

  To demonstrate the application of the functional constructed in
  \cref{ssec:functional-targeting-a-cat-state}, we consider a Kerr-nonlinear
  resonator with two-photon driving.
  In the rotating frame, it reads
  \begin{equation}
    \label{eq:h0-kerr}
    \op{H}_\mathrm{Kerr}/\hbar =
      -K \op{a}^\dagger\op{a}^\dagger\op{a}\op{a}
      + \varepsilon(t)\op{a}^2
      + \varepsilon^\ast(t)\big(\op{a}^\dagger\big)^2\,,
  \end{equation}
  where $K$ is the strength of the
  Kerr nonlinearity and $\varepsilon(t)$ is the complex-valued amplitude of the
  two-photon drive.
  This Hamiltonian has been realized experimentally via coupled Josephson
  junctions~\cite{LeghtasS15}. In this setting, various protocols exist for the
  generation of cat states using this~\cite{PurinQI17}
  or similar Hamiltonians~\cite{BartoloPRA16,MingantiSR16}.
  The set of reachable cat states in this system is limited by the two-photon
  drive in \cref{eq:h0-kerr} and depends on the initial state since the driving
  only allows for direct transfer between next-nearest energy levels, i.e., $\Delta n =\pm 2$.
  Therefore, when starting in the ground state $\ket{0}$, it is only possible to
  reach even-parity states.
  This means that only even cat states corresponding to a \spp{} $\varphi=0$
  are reachable.
  If the initial state is not an element
  of the even-parity subspace, then even cat states are unreachable,
  as transitions between the two parity subspaces are forbidden under two-photon driving.

  We compare the performance of a naive state-to-state optimization with our
  cat state optimization for the initial state
  $\ket{\psi_0}=\frac{1}{\sqrt{2}}\big(\!\ket{0} + \ket{1}\!\big)$.
  For the former optimization, we use the state-to-state functional,
  \begin{equation}
    \label{eq:s2s-functional}
    J_\mathrm{ss}\big(\psi(T)\big)
      = 1-\big|\!\braket{\psi(T)|\psi_\mathrm{tgt}}\!\big|\,.
  \end{equation}
  with the target state being an even cat state
  $\ket{\psi_\mathrm{tgt}}=\ket{\psi_\mathrm{cat}^+}$, similar to
  Ref.~\cite{RojanPRA14}.
  We present results for $\alpha=1.5$, but optimizations for
  different $\alpha$ yield similar results.
  Since the initial state $\ket{\psi_0}$ is not of even parity, it cannot evolve
  into an even cat state under two-photon driving.
  For the latter optimization, we use the cat state functional from
  \cref{ssec:functional-targeting-a-cat-state} with the cat term $J_\mathrm{cat}$
  chosen as in \cref{eq:j_cat_parity_original} targeting an arbitrary element from
  the cat state set with an arbitrary \spp{}.

  \begin{figure}
    \centering
    \includegraphics{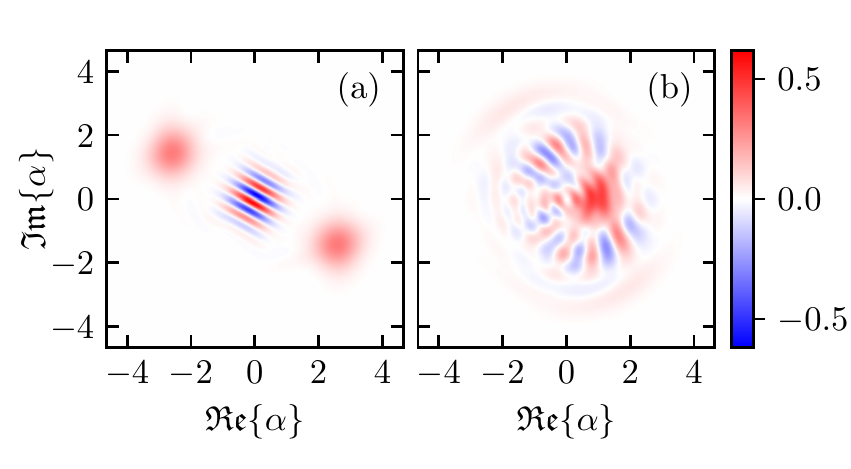}
    \caption{
      Comparison of the optimization results for two different functionals.
      The plot shows the Wigner distribution of the final states
      obtained with the cat state functional described in
      \cref{sec:framework-for-targeting-cat-states} in (a) and  with the
      state-to-state functional (\cref{eq:s2s-functional}) in (b).
    }
    \label{fig:2ph-comparison}
  \end{figure}
  \figureref{fig:2ph-comparison} shows the Wigner distribution of the final
  states after optimization with both functionals.
  While the optimization result for the cat state
  functional (\figref[(a)]{fig:2ph-comparison}) exhibits the classical phase
  space structure of a cat state, i.e., two coherent states with interference
  fringes in between them, the results of the state-to-state
  functional (\figref[(b)]{fig:2ph-comparison}) do not resemble a cat state.
  Furthermore, we observe that the optimized pulses we obtained via the cat state functional
  have a spectral width which does not exceed $400K$, where $K$ is again the strength of the Kerr nonlinearity.
  This confirms that we solidly stay in the regime where the rotating-frame
  expression in \cref{eq:h0-kerr} remains valid.
  In the original frame the frequency of the oscillator is usually in the
  order of $1000 K$, cf.~for example Ref.~\cite{GrimmN20}.
  This results in counter-rotating terms at frequencies around $4000K$, which is around one order
  of magnitude larger than the spectral width of our optimized pulses.

  To quantify the distance between the optimized states and the set of cat
  states we define the ``cat infidelity'' to be the smallest infidelity between
  the final state and any element of the cat state set,
  \begin{equation}
    \label{eq:cat_infidelity}
    I_\mathrm{cat}\big({\psi}\big)
      = \min_{\ket{\phi}\in\{\ket{{\psi}_{\mathrm{cat}}}\}}
        1-F({\phi}, {\psi})\,,
  \end{equation}
  where we use the overlap
  \begin{equation}
    \label{eq:fidelity}
    F({\phi}, {\psi}) = \big|\!\braket{\phi|\psi}\!\big|
  \end{equation}
  as state fidelity as defined in~\cite{Nielsen10} and $\{\ket{{\psi}_{\mathrm{cat}}}\}$ is the set of all
  cat states, which are described by \cref{eq:arb_cat_state}.
  Using this definition, we obtain a cat infidelity of $\approx 0.004$ for the
  optimization with the cat state functional and $\approx 0.235$ for the naive
  state-to-state functional approach.
  This confirms that the state obtained with the cat state functional is
  much closer to a cat state than the result achieved with the state-to-state
  functional.  This further elucidates, that targeting even cat states via
  a state-to-state functional for this Hamiltonian can only succeed when
  starting from an even initial state, as discussed earlier.

  In addition to the optimizations just presented, we also compared the
  performance of the state-to-state functional with the cat state functional for
  the initial state $\ket{\psi_0} =\ket{0}$.
  From this state, the target states defined for the state-to-state functionals
  are reachable with the two-photon drive added to the Kerr-Hamiltonian.
  As expected, we find that the amount of iterations required in the optimization algorithm
  depends on the chosen cat state functional $J_{\mathrm{cat}}$.
  While the cat state functional optimizing towards an even cat state, as
  defined in \cref{eq:j_cat_parity_original}, converges after around the same
  amount of iterations as the state to state functionals,
  $J_{\mathrm{cat},\varphi}$ (\cf \cref{eq:j_cat_arb_original}) takes
  about five times longer to converge.
  In case of the mixed parity initial state, we even need roughly another order of
  magnitude more iterations to reach convergence.
  This shows that more difficult optimization goals and more powerful optimization
  functionals often come at the price of higher numerical cost. However, since
  the state-to-state functional fails entirely for the mixed-parity initial state,
  this increased cost is well worth it.

  Our results illustrate that the success of a naive
  state-to-state optimization hinges critically on the choice of the
  target state and can completely fail if, e.g., the symmetry of
  the system is not properly considered.
  Although it would be possible to amend such issues by sampling the
  parameter space of cat states and perform optimizations until
  success is achieved for some set of parameters, such an approach would be
  numerically very expensive.
  In contrast, using the functional introduced in
  \cref{sec:framework-for-targeting-cat-states} allows for a maximally
  general optimization target and thus makes scanning of the parameter space
  completely obsolete.

%###############################################################################
%###############################################################################

\section{Optimization towards Entangled Cat States in a Bipartite System}
  \label{sec:optimization-towards-entangled-cat-states-in-a-bipartite-system}

  In this section we apply the functional from
  \cref{ssec:functional-targeting-an-entangled-cat-state,ssec:dissipation-adapted-functional}
  to a system consisting of a harmonic oscillator coupled to a qubit.
  One example for such a model is the dipolar transition between two circular
  Rydberg states interacting with a microwave cavity mode, which is realized in
  an experimental setup by Raimond, Haroche, Brune and
  co-workers~\cite{RaimondRMP01,Haroche06}.
  For simplicity, we model this system by a resonant
  Jaynes-Cummings-Hamiltonian in the interaction picture.
  After applying the rotating wave approximation the Hamiltonian is given by
  \begin{equation}
    \label{eq:h-jc}
    \op{H}_\mathrm{JC}/\hbar = g(
      \op{\sigma}_{+}\otimes\op{a} + \op{\sigma}_{-}\otimes\op{a}^\dagger
    ) + \varepsilon^\ast(t)\op{\sigma}_-\otimes\opopenone
      + \varepsilon(t)\op{\sigma}_+\otimes\opopenone\,,
  \end{equation}
  where $g$ describes the coupling strength between the harmonic oscillator and
  the qubit and $\varepsilon(t)$ is an external drive, which couples
  to the qubit and can be realized by \eg a microwave pulse.
  For our simulations, we use the parameters from the experimental setup
  in~\cite{RaimondRMP01,Haroche06}, where the coupling between the qubit and the
  cavity is given by $g=2\pi\cdot 50\,\mathrm{kHz}$, which is much smaller than
  the resonance frequency of the qubit and the cavity $\omega = 2\pi \cdot
  51\,\mathrm{GHz}$.
  The eigenstates of $\op{H}_\mathrm{JC}$ in \cref{eq:h-jc} read
  \begin{equation}
    \label{eq:jc-dressed-states}
    \ket{n, \pm} = \frac{1}{\sqrt{2}}\big(
      \ket{0}\otimes\ket{n+1}\pm\ket{1}\otimes\ket{n}
    \big)\,,
  \end{equation}
  with the corresponding eigenenergies $E^\pm_n = \pm \hbar g\sqrt{n}$.
  A more detailed introduction to the Jaynes-Cummings-Hamiltonian in the
  context of optimal control can be found, e.g., in Ref.~\cite{RojanPRA14}.

  To facilitate comparison between optimization results with different values of
  $|\alpha|$, we amend the functional from \cref{eq:functional_completely_entangled_state}
  by a third term $J_{|\alpha|}(\Psi)$ which
  allows to target cat states with a particular cat radius $|\alpha|$, called
  $|\alpha_\mathrm{tgt}|$ in the following.
  To accomplish this, we define a scalar cost function $f(x)$, which takes its
  minimal value at $x=|\alpha_\mathrm{tgt}|$.
  To compare the desired value $|\alpha_\mathrm{tgt}|$,
  with the actual cat radius $|\alpha|$ at final time $T$ we estimate $|\alpha|$
  from $\rho(T)$ by using the expression
  \begin{equation}
    \label{eq:alpha_final}
    |\alpha|^4 =
       \Tr{\Big[\big(\op{A}^\dagger\big)^2\op{A}^2\hat{\rho}(T)\Big]},
  \end{equation}
  where $\op{A}=\opopenone\otimes\op{a}$, as above.
  Strictly speaking, this expression only yields a sensible value for $|\alpha|$ if
  $\rho(T)$ is an entangled cat state, which is only true if the optimization
  is fully converged through minimization of the other functional terms.
  Still, the value for $|\alpha|$ obtained with \cref{eq:alpha_final} can be used
  as an estimate of the cat state radius, which in practice turns out to
  be a good approximation for states close to a cat state.
  For our calculations we used the functional
  \begin{equation}
    \label{eq:alpha_fix}
    J_{|\alpha|}(\Psi) = f\big(|\alpha|\big)
    =
    \frac{
      (|\alpha|^4-|\alpha_\mathrm{tgt}|^4)^2
    }{
      |\alpha_\mathrm{tgt}|^8
    }
    + \frac{(|\alpha|-|\alpha_\mathrm{tgt}|)^2}{|\alpha_\mathrm{tgt}|^2}
      \,.
  \end{equation}
  Our choice of the function $f$ proves to be particularly suitable since it possesses high
  gradients for both small and large arguments.
  This helps accelerating convergence during the optimization both far away and
  close to the target value.

  %%%%%%%%%%%%%%%%%%%%%%%%%%%%%%%%%%%%%%%%%%%%%%%%%%%%%%%%%%%%%%%%%%%%%%%%%%%%%%
  \subsection{Coherent Dynamics}
    \label{ssec:coherent-dynamics}
  %%%%%%%%%%%%%%%%%%%%%%%%%%%%%%%%%%%%%%%%%%%%%%%%%%%%%%%%%%%%%%%%%%%%%%%%%%%%%%

\begin{figure}
    \includegraphics[width=\columnwidth]{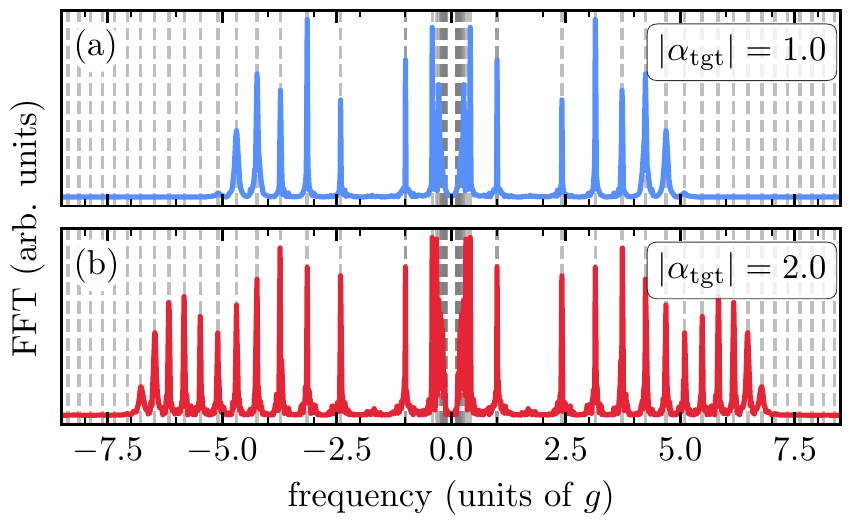}
    \caption{
      Spectra of the optimized pulses for optimization towards maximally entangled cat
      states with $|\alpha_\mathrm{tgt}|=1$ (top) and
      $|\alpha_\mathrm{tgt}|=2$ (bottom).
      The dashed lines indicate the transition frequencies between the
      eigenstates $\ket{n, \pm}$ of the drift Hamiltonian, defined in
      \cref{eq:jc-dressed-states}.
    }
    \label{fig:spectra-optims-compared}
  \end{figure}

  Using the functional $J_\mathrm{T}(\psi) = \varfunctional(\psi)
  + J_\mathrm{cat}(\psi) + J_{|\alpha|}(\psi)$
  we have performed numerical optimizations towards cat states with
  $|\alpha_\mathrm{tgt}|= 1$ and $|\alpha_\mathrm{tgt}|=2$, respectively, starting from the
  ground state of the atom-cavity system.
  For both values of $|\alpha|$ we have obtained states with cat infidelities of
  $I_\mathrm{cat}\big(\!\ket{\Psi(T)}\!\big)<10^{-3}$, where we use the same definition
  as in \cref{eq:cat_infidelity}, although this infidelity is now calculated with respect
  to the set of maximally entangled cat states defined in \cref{eq:ent_cat_state} to match
  the optimization goal.
  In \figref{fig:spectra-optims-compared}, we show the spectra of the optimized pulses.
  Both spectra exhibit sharp peaks located at the transition
  frequencies between the eigenstates of the atom-cavity system $\ket{n, \pm}$
  (\cf~\cref{eq:jc-dressed-states}) of the drift Hamiltonian
  $\op{\tilde{H}}_\mathrm{JC}$ with adjacent $n$, indicated by
  dashed lines in \figref{fig:spectra-optims-compared}.
  Since the qubit is driven, these are the only direct transitions
  between atom and cavity. In turn this allows for increasing or decreasing the number
  of excitations in the cavity by $\Delta n=\pm 1$ due to the interaction.
  It also explains why we observe spectral broadening when optimizing for cat
  states with larger $\alpha$. Such cat states require higher levels of the harmonic oscillator to be
  populated and thus higher-level transitions between the eigenstates of the
  atom-cavity system need to be driven by the pulse.
  The transition frequencies between $n$ and $n+1$ become either
  $\Delta\omega\propto \sqrt{n+1}+\sqrt{n}$ for
  $\ket{n,\pm}\rightarrow\ket{n+1,\pm}$, which we call type-(i) transitions
  or $\Delta\omega\propto\sqrt{n+1}-\sqrt{n}$ for
  $\ket{n,\pm}\rightarrow\ket{n+1,\mp}$, which we call type-(ii)
  transitions.
  Since the latter frequencies tend towards zero with larger $n$, the type-(ii) transition
  frequencies become more and more difficult to resolve for a given pulse
  duration.
  This only leaves the optimization algorithm to target the larger type-(i)
  transition frequencies to address higher-level transitions.
  Comparing the spectrum obtained for
  $|\alpha_\mathrm{tgt}|=1.0$ in \figref[(a)]{fig:spectra-optims-compared} and
  the spectrum for $|\alpha_\mathrm{tgt}|=2.0$ in
  \figref[(b)]{fig:spectra-optims-compared} confirms this interpretation.
  Indeed, the pulse in \figref[(b)]{fig:spectra-optims-compared} contains larger
  frequency components in comparison to the pulse in
  \figref[(a)]{fig:spectra-optims-compared} and exhibits a broader spectrum.

  We now inspect the optimized pulse obtained for
  $|\alpha_\mathrm{tgt}|=2.0$ in greater detail, noting that the optimization
  strategy obtained for $|\alpha_\mathrm{tgt}|=1.0$ turns out to be very similar.
  \begin{figure}
    \includegraphics[width=\columnwidth]{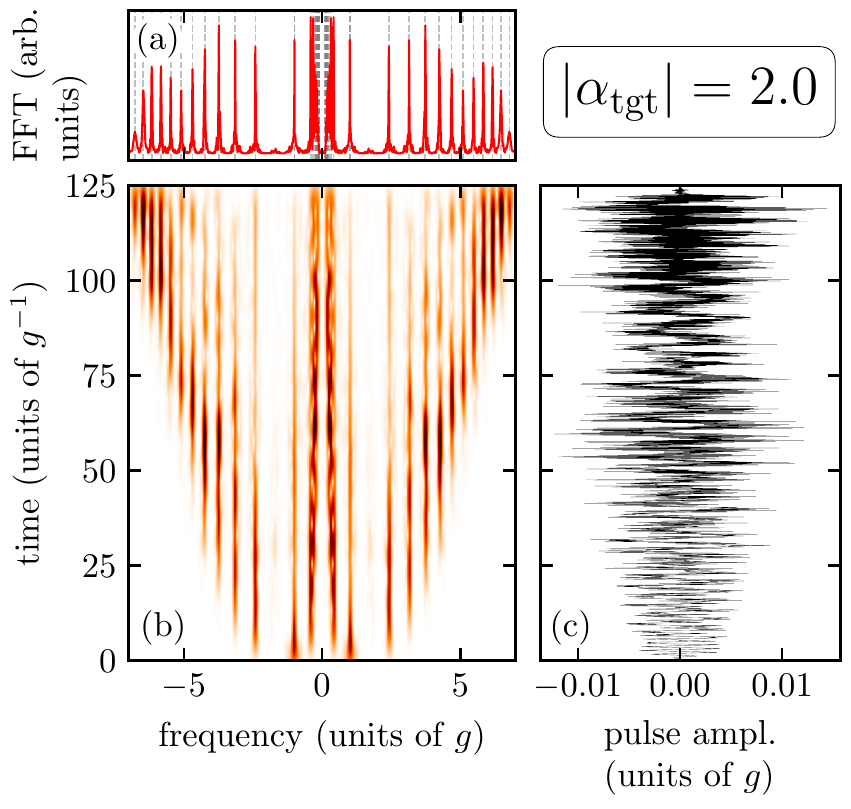}
    \caption{
      Detailed analysis of the optimized pulse towards
      $|\alpha_\mathrm{tgt}|=2.0$ showing the pulse in time domain ({c}),
      in frequency domain ({a}) and a time-frequency
      distribution ({b}) calculated via the Gabor transform.
      All quantities are expressed in units of the coupling strength $g$.
    }
    \label{fig:time_freq-spec-alpha20}
  \end{figure}
  \figureref[(c)]{fig:time_freq-spec-alpha20} depicts the real part of the
  optimized pulse in time domain, showing a consistent increase of
  the instantaneous pulse frequency with time.
  This behavior can be further elucidated with a Gabor transform of the pulse,
  \begin{equation}
    \label{eq:gabor-transform}
    G_\sigma(\tau, \omega) \propto
      \int_{\infty}^{\infty}
        %\frac{1}{\sqrt{2\pi}\sigma}
        \e{-\frac{(\tau-t)^2}{2\sigma^2}}
        \e{i\omega t} \varepsilon(t)
      \ \mathrm{d}t\,,
  \end{equation}
  shown in \figref[(b)]{fig:time_freq-spec-alpha20} with
  $\sigma=\frac{T}{4\sqrt{2\pi}}$.
  The Gabor transform reveals how the frequency components of the pulse change with time.
  For visual aid, the spectrum of the pulse from
  \figref[(b)]{fig:spectra-optims-compared} is plotted again in
  \figref[(a)]{fig:time_freq-spec-alpha20}.
  Indeed, the time-frequency distribution in
  \figref[(b)]{fig:time_freq-spec-alpha20} exhibits mainly contributions
  at the transition frequencies of the Jaynes-Cummings model.
  Additionally, the optimized pulse contains only a few frequency components at the
  beginning, with more - and in particular larger - frequencies added over time.
  This gradual driving of higher frequency components generates a population
  ascent towards higher levels, ultimately yielding the desired distribution.

  %---------------------------------------------------------------------------

  The pulse duration is an important property of an optimal control solution.
  It determines whether an operation can be carried out
  and influences the impact of dissipation. Specifically, finding the minimal time
  required to implement the physical target can help to
  limit the role of dissipative effects.
  Alternatively, dissipation can be included in the model - an approach that is
  followed in \cref{ssec:dissipative-dynamics}.
  Here, we restrict ourselves to coherent dynamics and use the functional
  developed in \cref{ssec:functional-targeting-an-entangled-cat-state} to
  investigate how fast cat states can be prepared for a Jaynes-Cummings
  Hamiltonian.
  The determination of the shortest time for generating
  or transforming states is an important task of optimal control theory. The resulting
  quantity is called the quantum speed limit~\cite{BhattacharyyaJPAMG83,MargolusPDNP98,CanevaPRL09,LevitinPRL09,GoerzPRA15},
  which is usually determined by a set of optimizations with
  varying pulse durations.
  The shortest pulse duration at which the objective can still be reliably reached
  is then used as an estimate for the quantum speed limit~\cite{CanevaPRL09,GoerzPRA15}.

\begin{figure}
    \centering
    \includegraphics{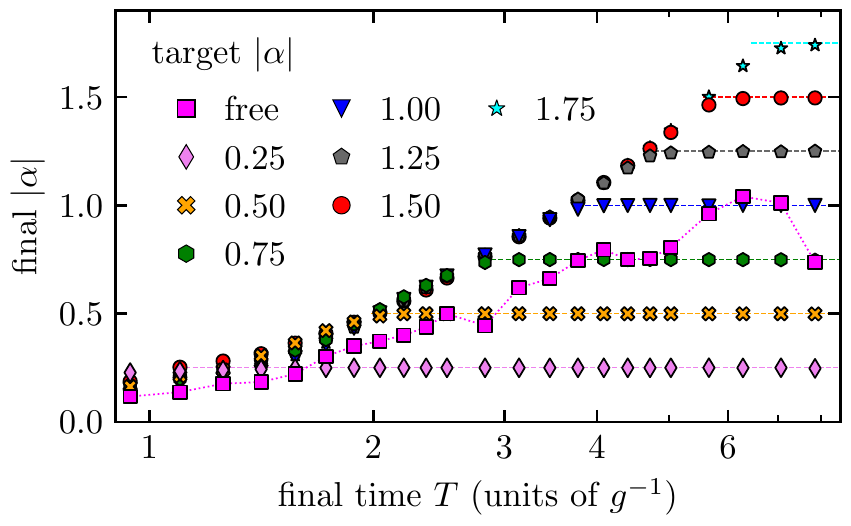}
    \caption{
      The final value of $|\alpha|$ plotted against the pulse duration $T$ for
      different optimizations.
      The different marker styles indicate optimizations performed with
      different target values $|\alpha_\mathrm{tgt}|$.
      The dashed auxiliary lines on the right, colored with the same color as
      the marker styles, are added to guide the eye.
    }
    \label{fig:qsl-alphas}
  \end{figure}

  Since the cat state functional is fairly complex, reaching convergence proved
  difficult in some of our calculations.
  Particularly the presence of plateaus in the optimization landscape posed an
  appreciable problem during some optimizations.
  These plateaus lead to asymptotically slow convergence if the value of $\lambda_a$, \cf \cref{eq:krotov_update_eq},
  is kept constant during the optimization.
  Hence, we have employed an additional line
  search inside Krotov's method to find the best value for the step size
  parameter $\lambda_a$ in each optimization step.
  With this modification, we have reliably obtained good convergence despite the functional's complexity.
  The results for the optimizations with different pulse durations are
  depicted in \figref{fig:qsl-alphas}.
  All data exhibit a similar trend, as the final value $|\alpha|$, \cf
  \cref{eq:alpha_final}, steadily increases for small values of $T$ and saturates
  at the desired value as soon as the pulse duration crosses a certain
  threshold.
  Once this threshold is reached, the optimization attains the desired value reliably.
  The smallest duration $T_\mathrm{QSL}^\alpha$ at which $|\alpha|$ reaches
  $|\alpha_\mathrm{tgt}|$ increases with the desired target value
  $|\alpha_\mathrm{tgt}|$.
  This suggests that $T_\mathrm{QSL}^\alpha$ indeed represents the quantum speed
  limit.
  The increase of $T_\mathrm{QSL}^\alpha$ with $|\alpha_\mathrm{tgt}|$
  can be directly justified with our previous findings in
  \figref{fig:time_freq-spec-alpha20}, showing that higher levels need to
  be sequentially populated and thus longer times are needed for higher
  excitation of the cat state.

  We have also performed optimizations towards entangled cat states without prescribing
  a target value for $|\alpha|$, i.e., without the term $J_{|\alpha|}$.
  The obtained ``free'' optimization results are plotted as magenta squares in
  \figref{fig:qsl-alphas}.
  As expected, the final values of the cat state radii $|\alpha|$ stay
  below all curves indicating the quantum speed limit for a given value of $|\alpha|$.
  However, the final values $|\alpha|$ do not always increase with larger
  pulse durations for the ``free'' optimization,
  whereas the quantum speed limit continuously increases towards larger
  values of $|\alpha_\mathrm{tgt}|$.
  The latter is in accordance with our previous findings in
  \figref{fig:time_freq-spec-alpha20}, showing that higher levels need to
  be populated one after another.
  Since larger excitations of the cat state, i.e., larger $|\alpha|$, require more
  transitions to higher levels, the larger time required to reach the target is
  not surprising.

  %%%%%%%%%%%%%%%%%%%%%%%%%%%%%%%%%%%%%%%%%%%%%%%%%%%%%%%%%%%%%%%%%%%%%%%%%%%%%%
  \subsection{Dissipative Dynamics}
    \label{ssec:dissipative-dynamics}
  %%%%%%%%%%%%%%%%%%%%%%%%%%%%%%%%%%%%%%%%%%%%%%%%%%%%%%%%%%%%%%%%%%%%%%%%%%%%%%

  Dissipation is expected to influence the preparation process of the entangled
  cat states.
  Including this influence in the optimization framework may allow for
  identifying control strategies that are better adapted to the presence of
  dissipation than those obtained with a coherent model.
  In the following, we first investigate the influence of dissipation on the
  coherently optimized results and then reoptimize the latter to find strategies
  better suited for dissipative dynamics.
  To this end, we consider an open quantum system described by a
  Gorini-Kossakowski-Sudarshan-Lindblad master equation~\cite{Breuer02} with
  $T_1$ relaxation of the HO,
  \begin{align}
    \label{eq:gksl}
    \frac{\mathrm{d}}{\mathrm{d}t} \op{\rho}(t)
      &= %\mathcal{L}\, \op{\rho}(t) \notag\\      &\equiv
      -\frac{i}{\hbar} \big[\op{H}(t), \op{\rho}(t)\big]
      + \kappa \Big(
        \op{L} \op{\rho} \op{L}^\dagger
        - \frac{1}{2}\big\{\op{L}^\dagger\op{L}, \op{\rho}(t)\big\}
      \Big)\,,
  \end{align}
  where $\op{\rho}$ is the joint density operator of HO and qubit and
  $\op{L}=\op{\openone}\otimes\op{a}$ with decay rate
  $\kappa$.

\begin{figure}
    \includegraphics[width=\columnwidth]{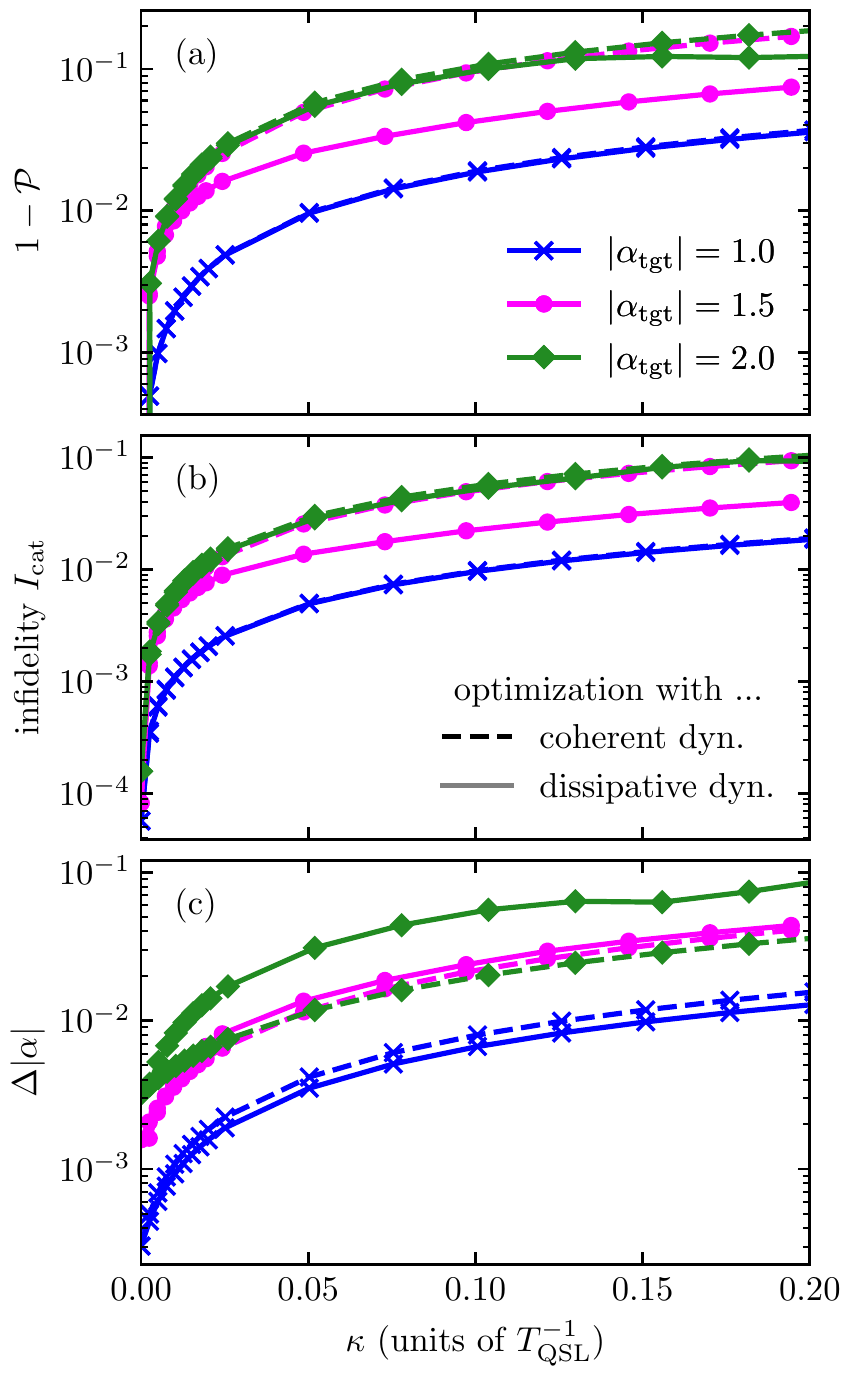}
    \caption{
      Dependence of the final state errors on the dissipation
      strength $\kappa$ for different target values
      $|\alpha_\mathrm{tgt}|$: (a) purity error, (b) cat infidelity, and (c) cat radius error.
      The solid lines correspond to pulses optimized with dissipation taken into account whereas
      the dashed lines display the results obtained using the coherently
      optimized pulses, but propagated with the corresponding dissipation
      strength.
    }
    \label{fig:functional-quantity-error}
  \end{figure}

  In order to analyze how the optimization protocols derived in the previous
  section are affected by dissipation, we inspect three quantities, or
  ``errors'', characterizing the final state and its quality, for three
  different target values $|\alpha_\mathrm{tgt}|$ in
  \figref{fig:functional-quantity-error}.
  The first quantity is the deviation of the
  purity $\mathcal{P}$ from that of a pure state, which is equivalent to the
  linear entropy defined in \cref{eq:lin_entropy}.
  Second, in order to quantify how close the final state is to the set of target
  states $\big\{\!\ket{{\Psi}_{\mathrm{cat}}}\!\big\}$, we define a ``cat
  infidelity'',
  \begin{equation}
    \label{eq:cat_infidelity_dm}
    I_\mathrm{cat}(\op{\rho})
      = \min_{\ket{\Psi}\in\{\ket{{\Psi}_{\mathrm{cat}}}\}}
        1-F\big(\op{\rho}, \ket{\Psi}\bra{\Psi}\big),
  \end{equation}
  where we use the generalized overlap
  \begin{equation}
    \label{eq:fidelity_dm}
    F(\op{\sigma}, \op{\rho})
      = \Tr{\smash[b]{\sqrt{\sqrt{\op{\rho}}\,\op{\sigma}\sqrt{\op{\rho}}}}}
  \end{equation}
  as state fidelity as defined in~\cite{Nielsen10}.
  The third quantity is the deviation of $|\alpha|$ from the
  target value $|\alpha_\mathrm{tgt}|$,
  \begin{equation}
    \label{eq:delta_alpha}
    \Delta|\alpha| = \big||\alpha|-|\alpha_\mathrm{tgt}|\big|.
  \end{equation}
  The target value $|\alpha_\mathrm{tgt}|$ determines the minimum time needed to prepare the cat state, the so-called quantum speed limit.
  We therefore discuss  the influence of dissipation by expressing the dissipation strength in units of the quantum speed limit $T_\mathrm{QSL}$.
  Following the same logic, the pulse duration $T$ is chosen to be
  approximately twice the quantum speed limit.
  The specific values are $T = 2.4\pi/g$ for $|\alpha_\mathrm{tgt}|=1.0$, $T= 3.5\pi/g$
  for $|\alpha_\mathrm{tgt}| = 1.5$, and $T= 5\pi/g$ for $|\alpha_\mathrm{tgt}|=
  2.0$.

  %----------------------------------------------------------------------------
  \figureref[({a})]{fig:functional-quantity-error} depicts the deviation of the
  final state purity from the ideal purity with growing dissipation strength.
  The dashed lines correspond to the performance of pulses optimized for
  coherent evolution but evaluated in the presence of dissipation.
  The purity errors show a qualitatively similar behavior for different
  $|\alpha_\mathrm{tgt}|$ and, as expected, increase with dissipation strength.
  The solid lines display the results of reoptimization in the presence of
  dissipation using the pulses obtained without dissipation as guess pulse.
  They follow a similar trend as the coherently optimized results.
  Upon closer inspection, however, they exhibit improvements, which are most
  pronounced for $|\alpha_\mathrm{tgt}|=1.5$ (magenta lines).
  We analyze these improvements in more detail below.

  Dissipation does not only affect the purity, but we also find a larger
  infidelity of the final states.
  The cat infidelity as defined in \cref{eq:cat_infidelity_dm} is plotted in
  \figref[(b)]{fig:functional-quantity-error}.
  Analogously to \figref[(a)]{fig:functional-quantity-error}, infidelity
  and dissipation strength are correlated and the shapes of the curves
  resemble those depicted in \figref[(a)]{fig:functional-quantity-error}.
  Finally, \figref[(c)]{fig:functional-quantity-error} analyzes
  the error of $|\alpha|$ with respect to the target value
  $|\alpha_\mathrm{tgt}|$ as a function of the decay rate $\kappa$.
  Since the model accounts for decay of the harmonic oscillator,
  $\alpha$ is particularly affected by dissipation,
  reducing $\alpha$ with increasing $\kappa$.
  Indeed, the curves in \figref[(c)]{fig:functional-quantity-error}
  show this behavior and follow a similar trend as in
  \figref[(a-b)]{fig:functional-quantity-error} for both the coherently and the
  reoptimized pulses.
  Note that the reoptimization does not reduce the final-time cat radius error for the
  two larger values of $|\alpha_\mathrm{tgt}|$, $|\alpha_\mathrm{tgt}|=1.5$ (magenta)
  and $|\alpha_\mathrm{tgt}|=2.0$ (green). This is not surprising for the following
  reason. The optimization targets a sum of  terms which are balanced against each other.
  Since the function $f$ used in the definition of $J_{|\alpha|}$
  is comparatively insensitive to errors in the desired value of $|\alpha|$, the
  value of $|\alpha|$ can slightly deteriorate in favor of improvements in the
  other terms. If needed, this could be counteracted by changing the weight of
  $J_{|\alpha|}$ or even changing the function $f$ used when defining
  $J_{|\alpha|}$.

  %----------------------------------------------------------------------------
  \begin{figure}
    \includegraphics[width=\columnwidth]{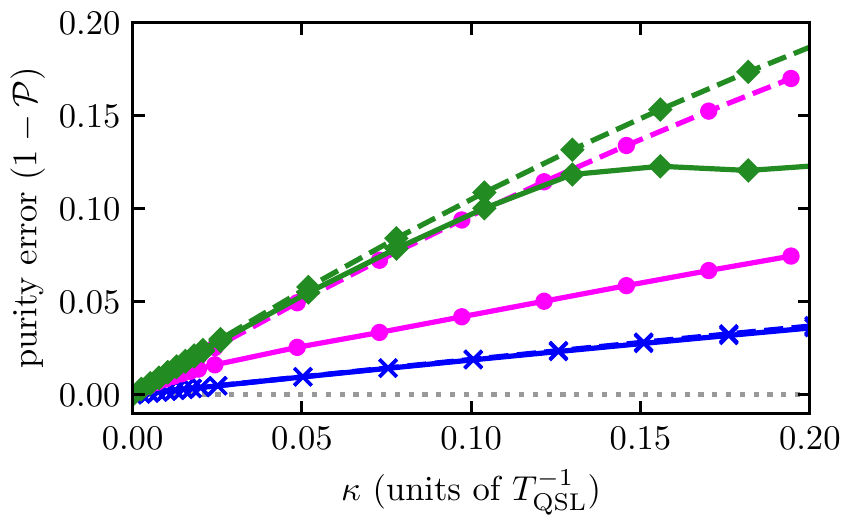}
    \caption{
      The same data as in \figref[({a})]{fig:functional-quantity-error} but with
      linear scale for the purity to illustrate the different
      improvements under reoptimization with dissipation (solid) compared
      to coherently optimized results (dashed).
      Same line styles and color code as in
      \figref{fig:functional-quantity-error}.
    }
    \label{fig:purity}
  \end{figure}
  Despite the similar behavior of the three quantities presented in
  \figref{fig:functional-quantity-error} on first glance, a more detailed analysis of the curves
  reveals three different types of adjustments that the reoptimization introduces.
  We illustrate these in \figref{fig:purity} by changing the scale of
  plot \figref[(a)]{fig:functional-quantity-error} to highlight the differences
  between the coherently optimized (dashed) and reoptimized (solid) results.
  The blue curve, corresponding to $|\alpha_\mathrm{tgt}|=1.0$, shows no
  improvement due to the reoptimization, which indicates that the solution
  found without account of dissipation is already robust.
  This is corroborated by the observation that the other quantities in
  \figref[(a,c)]{fig:functional-quantity-error} do not significantly improve
  either.
  In contrast, the purity of the reoptimized result for
  $|\alpha_\mathrm{tgt}|=1.5$ (magenta) is strongly enhanced compared to the
  original result for all decay rates $\kappa$ considered.
  An improvement under reoptimization is also observed for
  $|\alpha_\mathrm{tgt}|=2.0$ (green), except for small values of $\kappa$.
  The improvement in both cases suggests that the strategies found by the
  coherent optimization are not optimal once dissipation is taken into account
  and indicates the importance of considering dissipation in the optimization for
  obtaining more robust strategies.
  We also studied the generation of cat states with larger values of
  $\alpha$, which revealed a similar behavior, albeit with a much stronger
  purity loss.
  For very large $\alpha$ this loss of purity becomes so strong that
  the optimization fails which illustrates the practical limitation that is imposed by the
  decay on exciting the system coherently.

  %----------------------------------------------------------------------------
  \begin{figure*}
    \includegraphics[width=\textwidth]{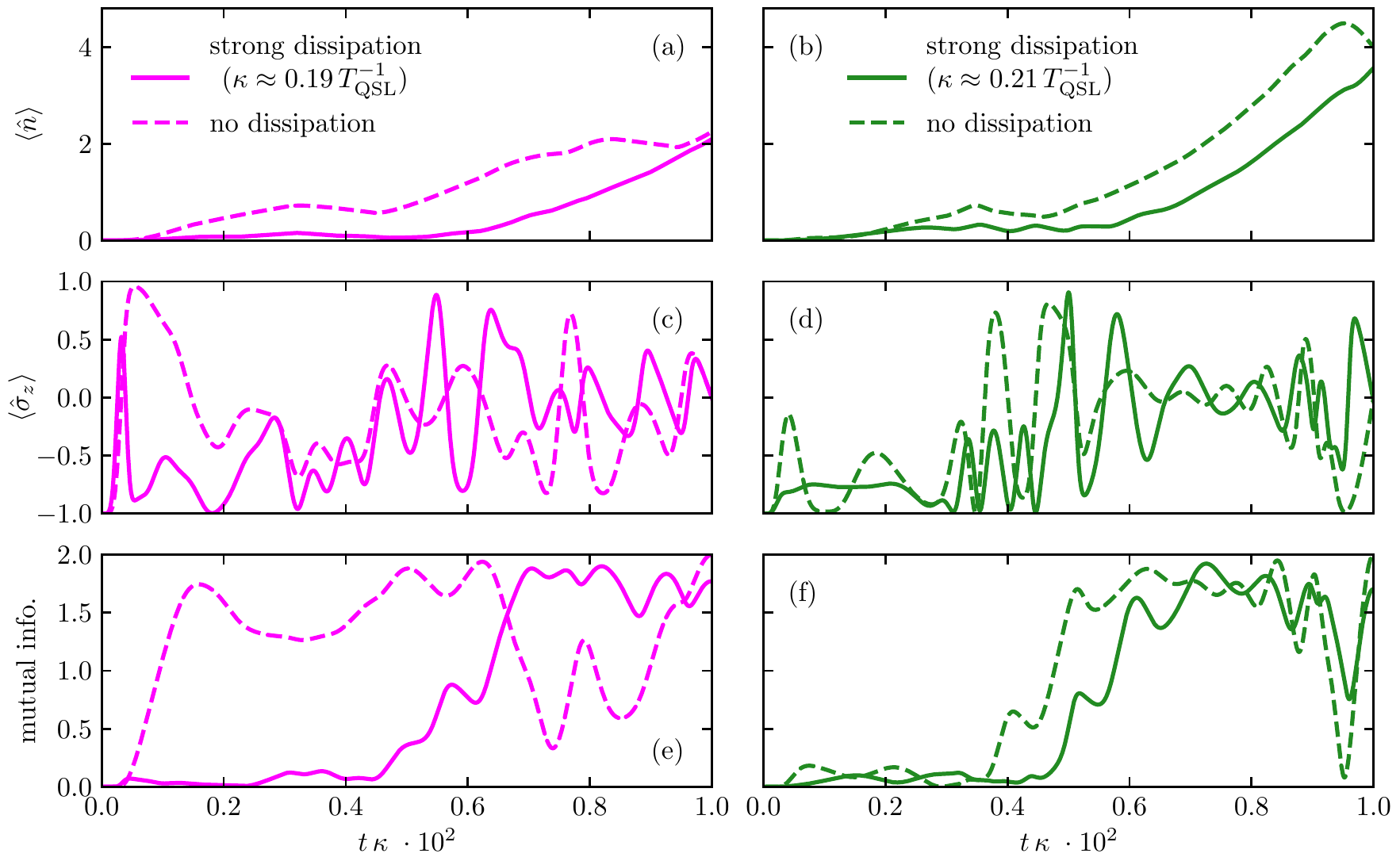}
    \caption{
      Comparison of different strategies:
      The solid lines depict the dynamics induced by the pulse optimized with
      strong dissipation and the dashed lines depict the dynamics with the
      coherently optimized pulses.
      The left and right column correspond to $|\alpha_\mathrm{tgt}|=1.5$ and
      $|\alpha_\mathrm{tgt}|=2.0$, respectively.
      The panels (a-b) and (c-d) show the average excitation of the harmonic
      oscillator, \cref{eq:expn}, and of the qubit,
      \cref{eq:expsigmaz}, respectively, as well as the mutual information in (e-f),
      \cref{eq:vn_mutual_info}.
    }
    \label{fig:system-quantities}
  \end{figure*}
  We investigate the change in control strategy next.
  To this end, we focus on the results for $|\alpha_\mathrm{tgt}|=1.5$ and
  $|\alpha_\mathrm{tgt}|=2.0$, showing signs of a strategy change as
  discussed above.
  To analyze the dynamics of the harmonic oscillator, the mean number of
  excitations,
  \begin{equation}
    \label{eq:expn}
    \braket{\op{n}}
    \equiv \Tr\big[(\op{\openone}\otimes \op{n})\op{\rho}(t)\big],
  \end{equation}
  is plotted over time in \figref[(a-b)]{fig:system-quantities}.
  Analogously \figref[(c-d)]{fig:system-quantities} shows the average excitation
  of the qubit,
  \begin{equation}
    \label{eq:expsigmaz}
    \braket{\op{\sigma}_z}
      \equiv \Tr\big[(\op{\sigma}_z \otimes \op{\openone})\op{\rho}(t)\big],
  \end{equation}
  as a function of time.
  Additionally, the interaction between the subsystems is analyzed in
  \figref[(e-f)]{fig:system-quantities} by means of the mutual information,
  defined in \cref{eq:vn_mutual_info}.
  We start by describing the dynamics for the case $|\alpha_\mathrm{tgt}|=1.5$.
  In \figref[(a)]{fig:system-quantities}, the dynamics optimized under pulses
  with and without dissipation show clear differences.
  Under the coherently optimized pulses (dashed line), the HO gets excited
  already shortly after the beginning.
  In contrast, the pulse optimized taking dissipation into account (solid line)
  keeps the excitation in the harmonic oscillator very small until $t\approx
  T/2$, where it starts to increase linearly to
  $\braket{\op{n}}\approx|\alpha|^2$.
  Overall, the excitation induced by the pulse optimized without dissipation
  is above the one for the reoptimized pulse at all times.
  The strategy of reducing excitation is not surprising as more
  excitation in the harmonic oscillator is directly related to a stronger decay.
  The excitation dynamics on the qubit, depicted in
  \figref[(c)]{fig:system-quantities}, is oscillatory and does not
  immediately reveal an underlying strategy.
  As we will see below, it is important that, for strong dissipation,
  $\braket{\op{\sigma}_z}$ oscillates around zero in the last third of the
  time interval.
  \figref[(e)]{fig:system-quantities} depicts the mutual information and thus
  the correlations between the subsystems.
  Note, that in case of the coherent dynamics, the mutual information indicates
  the entanglement between the systems, while for the mixed case the
  interpretation of the mutual information is more intricate.
  While in the coherently optimized case the harmonic oscillator and qubit
  become strongly entangled from the beginning on, the strategy for the
  reoptimized pulses is to keep the subsystems independent from each other until
  about $t\approx T/2$, and then continue with strong correlations between the
  subsystems.
  During this phase of strong correlations, excitation is directly transferred
  to the harmonic oscillator, which explains the straight excitation increase
  in \figref[(a)]{fig:system-quantities}.
  All in all, the strategy identified by the optimization algorithm in the
  presence of dissipation is to wait in the beginning and then generate the cat
  state as fast as possible in the end.
  This simply reduces the time during which excitation in the harmonic
  oscillator is exposed to decay, yielding higher quality final states.

  In case of $|\alpha_\mathrm{tgt}|=2.0$, the strategy change is more
  subtle.
  For both curves in \figref[(b)]{fig:system-quantities}, $\braket{\op{n}}$ is
  almost constant around zero in the beginning and linearly grows to the desired
  value of $\braket{\op{n}}_\mathrm{tgt}=|\alpha_\mathrm{tgt}|^2 =4.0$ after
  about half of the time.
  The difference between both curves is the final peak of the dynamics under the
  coherently optimized pulse (dashed line), which surpasses the desired value
  just before the end and finally decreases to match the desired excitation.
  In contrast, the solid curve, representing the dynamics under the reoptimized
  pulse, approaches the final value in an almost straight line, a behavior
  already observed for the reoptimized dynamics in
  \figref[(a)]{fig:system-quantities}.
  The dynamics of $\big\langle\op{\sigma}_z\big\rangle$, depicted in
  \figref[(d)]{fig:system-quantities}, is again more complex.
  The first part is clearly dominated by small fluctuations around the
  ground state of the qubit.
  Around half time, %When transitioning to the second half,
  both dynamics exhibit strong
  oscillations before $\braket{\op{\sigma}_z}$ remains close to zero in
  the interval between $t\approx 0.65\cdot 10^{-2}\,\kappa^{-1}$ and $t\approx
  0.85\cdot 10^{-2}\,\kappa^{-1}$.
  Thereafter, the oscillations grow stronger again, ultimately ending close to
  the desired value of $\braket{\op{\sigma}_z}= 0$.
  Finally, the dynamics of the mutual information in
  \figref[(f)]{fig:system-quantities} are similar to each other and those found
  for the dissipatively optimized dynamics in
  \figref[(e)]{fig:system-quantities}.
  For the pulse obtained under strong dissipation, the mutual
  information increases a bit later compared to the coherently optimized case.
  Also, the mutual information for $|\alpha_\mathrm{tgt}|=2.0$ exhibits
  a significant dip just before the end, which indicates that the system briefly goes to
  a relatively uncorrelated state just before the entangled cat state is created.
  Apart from this detail, both strategies for $|\alpha_\mathrm{tgt}|=2.0$
  follow a similar structure as the reoptimized dynamics discussed in
  \figref[(e)]{fig:system-quantities}.
  Despite this similarity of the dynamics, we observe a strong improvement in purity and even
  fidelity for large dissipation.
  The mutual information of the reoptimized pulse in
  \figref[(e)]{fig:system-quantities} also exhibits a speedup, compared to the
  coherently optimized dynamics, which is reflected in the delayed rise of the
  solid curve, compared to the dashed one.
  However, since the speedup is only marginal, the enhancement must also be
  related to avoiding the final peak in the average excitation of the harmonic
  oscillator.

  %----------------------------------------------------------------------------
  \begin{figure}
    \includegraphics[width=\columnwidth]{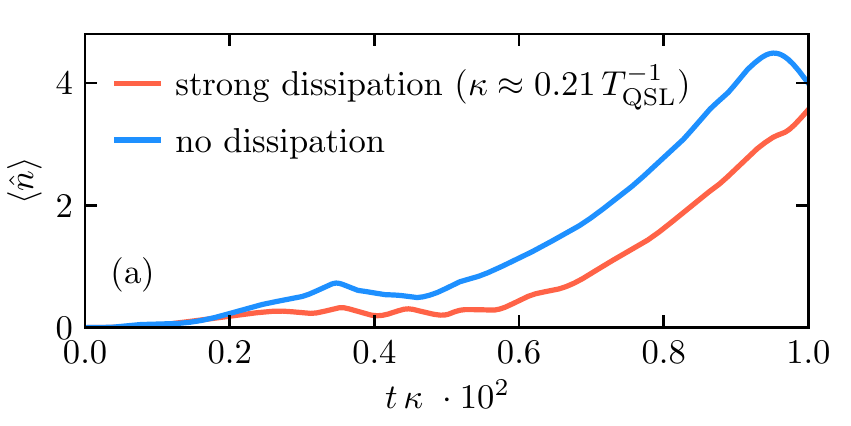}
    \includegraphics[width=.49\columnwidth]{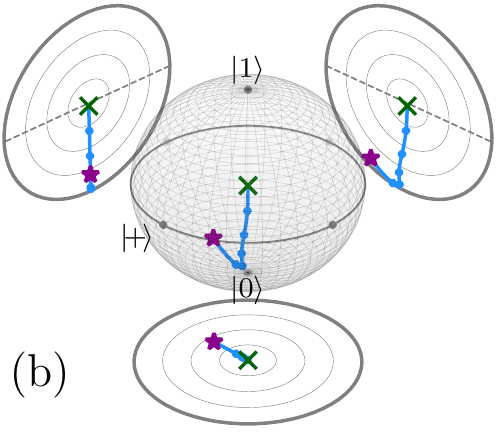}\hfill
    \includegraphics[width=.49\columnwidth]{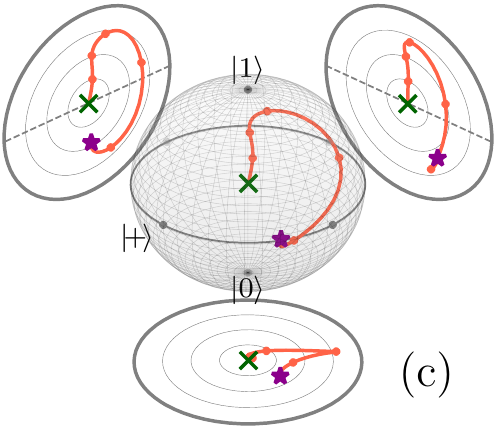}
    \caption{
      Visualization of the strategy change, storing excitation in the qubit
      instead of the harmonic oscillator, when optimizing with strong
      dissipation.
      Panel (a) is the same plot as in
      \figref[(b)]{fig:system-quantities} and panels
      (b-c) show the dynamics of the qubit in Bloch sphere
      representation.
      For the sake of clarity, we only show the final part of the dynamics in
      (b-c), as indicated by the gray dashed lines in (a).
    }
    \label{fig:oct-strategies}
  \end{figure}
  To further illustrate the strategy change for $|\alpha_\mathrm{tgt}|=2.0$,
  the qubit dynamics close to the final peak is presented in
  \figref[(b-c)]{fig:oct-strategies} together with the data from
  \figref[(b)]{fig:system-quantities} replotted in
  \figref[(a)]{fig:oct-strategies} for clarity.
  Here, the star and cross in \figref[(a)]{fig:oct-strategies} indicate the time
  period plotted in \figref[(b-c)]{fig:oct-strategies}.
  For the optimization not taking into account dissipation
  (\figref[(b)]{fig:oct-strategies}), the qubit starts by evolving to the
  ground state $\ket{0}$ becoming disentangled from the harmonic oscillator as
  already indicated by the dip in the mutual information in
  \figref[(e)]{fig:system-quantities}.
  It continues and evolves to the maximally entangled state at the center of the
  Bloch sphere.
  For the reoptimized dynamics (\figref[(c)]{fig:oct-strategies}),
  the qubit instead starts to evolve towards the excited state
  $\ket{1}$, but does not completely reach it, due to the purity
  reduction caused by dissipation.
  From there it evolves to the center of the Bloch sphere, again in an almost
  straight line.
  We thus find that the two strategies approach the
  final state in a similar way, yet from different sides of the Bloch sphere.
  In the case of the coherently optimized pulse (blue), this means that for
  the final part of the protocol, excitation is transferred from the harmonic
  oscillator to the qubit.
  This coincides with the observation of the peak around $t=0.95\cdot 10^{-2}\,\kappa^{-1}$ in
  \figref[(a)]{fig:oct-strategies}.
  It represents the excitation excess that is stored in the HO and later
  transferred to the qubit.
  Since the dissipation punishes more excitation in the harmonic oscillator, the
  strategy for the reoptimized pulse is to instead store excitation in the
  qubit and transfer the surplus excitation to the harmonic
  oscillator in the end, thus protecting it from decay as long as possible.

%###############################################################################
%###############################################################################

\section{Conclusions}
  \label{sec:conclusions}

  We have developed an optimization framework to target
  cat states and entangled cat states in bipartite systems.
  The corresponding functionals target the
  whole set of cat states instead of individual elements
  from this set. As a result, they provide the maximal flexibility for the
  optimization algorithm to steer the system towards the most suitable state from
  the set in a given physical setup.
  The composite functionals consist of several terms, which separately
  check whether the state is an eigenstate of the annihilation operator, whether
  it is an equally weighted superposition, and, in case of a bipartite entangled cat,
  whether the two subsystems are maximally entangled. Moreover, we have shown how to
  adjust the functionals to specify particular cat state properties, such as
  a desired displacement of the coherent states constituting the cat
  superposition.

  We have performed example optimizations using these functionals for
  a Kerr-nonlinear oscillator with two-photon driving.
  By directly comparing their performance with a naive state-to-state approach, we have been
  able to show that our framework reliably finds cat states, whereas the
  state-to-state approach needs to be manually adjusted to a reachable state in
  advance.
  This demonstrates the power of the cat state functionals which allow to
  avoid analyzing the reachable set of cat states in advance.
  Furthermore, we have successfully applied the functionals for
  maximally entangled bipartite cat states in an archetypical Jaynes-Cummings model.
  We have analyzed the structure of the optimized pulses which show a consecutive
  transfer of population towards higher energy levels as control mechanism.
  Moreover, we have investigated how fast cat
  states can be prepared and determined the quantum speed limit which we found to be directly
  connected to the excitation of the cat states, i.e., to the cat radius.

  Finally, we adopted our framework to open quantum systems and investigated the influence of dissipation
  in the Jaynes-Cummings model, comparing the performance of coherently optimized pulses
  with those optimized in the presence of decay of the harmonic oscillator.
  Accounting for the decay during the optimization has allowed us to improve the final state
  fidelity for states with large cat radius which are most affected by the decay.

  Inspecting the dynamics under the optimized pulses, we have been able to identify the strategy
  changes obtained when taking dissipation into account.
  In the presence of decay, it is more advantageous to keep excitation of the harmonic
  oscillator low as long as possible.
  While this by itself is not surprising, the optimization identifies the most suitable protocol
  depending on the desired cat radius.
  For sufficiently small radii, the cat is simply generated as fast as possible towards the end of the protocol.
  For larger cat radii, the best way to keep the oscillator excitation low is by storing
  excitation temporarily in the qubit.
  These results illustrate how taking dissipation into account during the optimization of a desired
  quantum process allows for identifying control strategies which are more robust than those obtained by optimizing coherent dynamics.

  In the future our functionals could be further adapted to target
  more complex sets of states, for example, multicomponent
  cat states~\cite{ZurekN01,DalvitNJP06,ToscanoPRA06,BergmannPRA16} or
  (entangled) multipartite cat
  states~\cite{MunroPRA02,GilchristJOBQSO04,WangSA22}.
  Another potential future avenue of our work is to extend the
  framework to the generation of squeezed states or
  squeezed cat states, which have recently attracted interest in
  quantum error correcting codes~\cite{SchlegelPRA22}.
  It will also be interesting to apply the cat state optimization framework derived
  here to open quantum systems that require a description of  the environment's influence
  beyond a phenomenological decay, in particular to systems with non-Markovian dynamics.
  For example, superconducting circuits are subject to $1/f$ noise~\cite{PaladinoRMP14} which results in non-Markovian dynamics.
  Beyond identifying strategies that are best adapted to the open system properties, this may allow
  for exploiting non-Markovianity as a resource for
  control, see e.g.~\cite{ReichSR15,KochJPCM16} and references therein.
  The present results thus increase the utility of the quantum optimal control
  toolbox~\cite{GlaserEPJD15,KochEQT22} for practical applications in the quantum technologies.

\begin{acknowledgments}
  Financial support from the federal state of Hesse, Germany via the SMolBits
  project within the LOEWE program is gratefully acknowledged.
\end{acknowledgments}

%###############################################################################
%###############################################################################
\newpage
\appendix

\section{Derivation of the cat term for arbitrary {\spp}s}
  \label{app:derivation-spp-functional}

  We show that an eigenstate of $\op{a}^2$ (\cf
  \cref{eq:general_coh_suppos}) is of the form of
  \cref{eq:arb_cat_state} if and only if
  \begin{equation}
    \label{eq:cat_condition_app}
    \big\lVert \bar{a}\Ket{\psi_{a^2}} - \op{a}\Ket{\psi_{a^2}} \big\rVert^2 =
    \big\lVert \bar{a}\Ket{\psi_{a^2}} + \op{a}\Ket{\psi_{a^2}} \big\rVert^2,
  \end{equation}
  where $\bar{a} \equiv \sqrt{\Braket{\psi_{a^2}|\op{a}^2|\psi_{a^2}}}$, which
  becomes $\bar{a}=\sqrt{\alpha^2}=\alpha$ if the state is an eigenstate of
  $\op{a}^2$.
  Expanding both sides of the equation yields
  \begin{align*}
    \bar{a}\Ket{\psi_{a^2}} &- \op{a}\Ket{\psi_{a^2}} =
      \\
        &= c_1\alpha\ket{\alpha} + c_2\alpha\ket{-\alpha}
        -c_1\alpha\ket{\alpha} + c_2\alpha\ket{-\alpha}
      \\
        &=2c_2\alpha\ket{-\alpha},
      \\
    \bar{a}\Ket{\psi_{a^2}} &+ \op{a}\Ket{\psi_{a^2}} =
      \\
        &= c_1\alpha\ket{\alpha} + c_2\alpha\ket{-\alpha}
        +c_1\alpha\ket{\alpha} - c_2\alpha\ket{-\alpha}
      \\
        &=2c_1\alpha\ket{\alpha}.
  \end{align*}
  This implies
  \begin{align*}
    \big\lVert \bar{a}\Ket{\psi_{a^2}} - \op{a}\Ket{\psi_{a^2}} \big\rVert^2
      &= 4|c_2|^2|\alpha|^2, \\
    \big\lVert \bar{a}\Ket{\psi_{a^2}} + \op{a}\Ket{\psi_{a^2}} \big\rVert^2
      &= 4|c_1|^2|\alpha|^2,
  \end{align*}
  such that \cref{eq:cat_condition_app} reduces to
  \begin{equation*}
    4|c_2|^2|\alpha|^2 = 4|c_1|^2|\alpha|^2
    \qquad
    \Longleftrightarrow
    \qquad
    |c_2|^2 = |c_1|^2
  \end{equation*}
  as desired.
  By calculating the state norms, \cref{eq:cat_condition_app} can be further
  simplified to
  \begin{align}
    \label{eq:norms_evaluated}
    \big\lVert \bar{a}\Ket{\psi_{a^2}} \pm \op{a}\Ket{\psi_{a^2}} \big\rVert^2=&
      \notag\\
      =& |\bar{a}|^2
        \pm \bar{a}^\ast \braket{\psi_{a^2}|\op{a}|\psi_{a^2}}
      \notag\\
      &\pm \bar{a} \braket{\psi_{a^2}|\op{a}^\dag|\psi_{a^2}}
        + \braket{\psi_{a^2}|\op{a}^\dag\op{a}|\psi_{a^2}}
  \end{align}
  which yields the condition
  \begin{equation}
    \label{eq:arb_phase_condition_app}
    \bar{a}^\ast \braket{\psi_{a^2}|\op{a}|\psi_{a^2}}
      + \bar{a}\braket{\psi_{a^2}|\op{a}^\dag|\psi_{a^2}}
      \overset{!}{=} 0,
  \end{equation}
  that is used for the cat term in \cref{eq:j_cat_arb_original}.

\section{Calculation of the costate}
  \label{app:calculation-co-state}
  In the following, we briefly summarize the calculation of the costates,
  \begin{equation}
    \ket{\chi} = -\nabla_{\bra{\psi}} J_T
      = -\frac{\partial J}{\partial\bra{\Psi}}\Big|_{t=T} \,.
  \end{equation}
  For the terms presented in
  \cref{sec:framework-for-targeting-cat-states}, most of the costates can be
  directly calculated by using the relation
  \begin{equation}
    \frac{\partial}{\partial\bra{\Psi}} \Tr(\op{O}\op{\rho})
    =\frac{\partial}{\partial\bra{\Psi}} \braket{\psi|\op{O}|\psi}
    = \op{O}\ket{\psi},
  \end{equation}
  where $\op{\rho} = \ket{\psi}\bra{\psi}$.
  However, the calculation of the costates for the cat term in
  \cref{eq:j_purity_arb_entangled_cats} is not as straightforward.
  To calculate the derivative, we first express
  \cref{eq:j_purity_arb_entangled_cats} as
  \begin{align}
    J_{\mathrm{P}}(\Psi) &\equiv
      J_{\mathrm{P}}(\Psi_1, \Psi_2)\Big|_{\Psi_1 = \Psi_2 = \Psi} \\
    &= 2\Tr(\op{\rho}_{1,\mathrm{HO}}\:\op{\rho}_{2,\mathrm{HO}})
      \Big|_{\op{\rho}_1 = \op{\rho}_2 = \op{\rho}} - 1
    \label{eq:purity_func_trick}
  \end{align}
  with $\op{\rho}_{i,\mathrm{HO}}=\trace[\mathrm{qubit}]{\op{\rho}_i}$.
  Using this relation, we calculate the costate of the cat term
  $\ket{\chi_{\mathrm{P}}}$ as
  %\begin{widetext}
    \begin{align}
      \ket{\chi_{\mathrm{P}}}
        &= -\frac{\partial J_{\mathrm{P}}}{\partial\bra{\Psi}}\Big|_{t=T}
         = -\bigg(\frac{\partial J_{\mathrm{P}}}{\partial\bra{\Psi_1}}
          +\frac{\partial J_{\mathrm{P}}}{\partial\bra{\Psi_2}}\bigg)
            \Bigg|_{\Psi_1=\Psi_2=\Psi}
      \notag\\
        &= -\bigg(
            \Big(
              2\frac{\partial}{\partial\bra{\Psi_1}}
              \Tr\big(
                (\op{\rho}_{2,\mathrm{HO}}\otimes\opopenone_\mathrm{qubit})
                \:\op{\rho}_{1}\big)
              \Big)
      \notag\\
        &\quad\
            +\Big(
              2\frac{\partial}{\partial\bra{\Psi_2}}
              \Tr\big(
                (\op{\rho}_{1,\mathrm{HO}}\otimes\opopenone_\mathrm{qubit})
                \:\op{\rho}_{2}\big)
              \Big)
          \bigg)
            \Bigg|_{\Psi_1=\Psi_2=\Psi}
      \notag\\
        &= -\Big(
          2\big(\op{\rho}_{2,\mathrm{HO}}\otimes\opopenone_\mathrm{qubit}\big)
              \ket{\Psi_1(T)}
    \notag\\
      &\quad\
          +2\big(\op{\rho}_{1,\mathrm{HO}}\otimes\opopenone_\mathrm{qubit}\big)
              \ket{\Psi_2(T)}
            \Big)\bigg|_{\Psi_1=\Psi_2=\Psi}
      \notag\\
        &= -4\big(\op{\rho}_\mathrm{HO}\otimes\opopenone_\mathrm{qubit}\big)
          \ket{\Psi(T)}.
    \end{align}
  %\end{widetext}

\section{Proof for the equivalence of \cref{eq:ent_cat_state}
    and \cref{eq:functional_completely_entangled_state}}
  \label{app:proof-ent-cat-equiv}

  In \cref{ssec:functional-targeting-an-entangled-cat-state} a functional is
  constructed which allows for optimizing towards states of the form given in
  \cref{eq:functional_completely_entangled_state}.
  At first glance, this appears to be a different form compared to the desired
  entangled cat states in \cref{eq:ent_cat_state}.
  In the following we show that both expressions are indeed equivalent.
  To this end, we first write the eigenstate of the annihilation operator
  $\op{a}^2$ defined in \cref{eq:general_coh_suppos} as
  \begin{equation}
    \ket{\psi_{a^2}} = c_0\ket{\alpha} + c_1\ket{-\alpha}
      = d_+ \ket{\psi_\mathrm{cat}^+} + d_- \ket{\psi_\mathrm{cat}^-},
  \end{equation}
  where $|d_+|^2 + |d_-|^2 = 1 \quad d_\pm\in\mathbb{C}$.
  Using this, we rewrite
  \cref{eq:general_state_arb_qubit_completely_entangled} as
  \begin{align}
    \ket{\Psi_{\mathrm{ent}}}
      &= \frac{1}{\sqrt{2}}\big(
          \ket{b_0} \otimes \ket{\psi_{\mathrm{0,a^2}}} +
          \ket{b_1} \otimes \ket{\psi_{\mathrm{1,a^2}}}
        \big)
    \notag\\
      &= \frac{1}{\sqrt{2}}\Big(
          \ket{b_0} \otimes \big(d_{0+} \ket{\psi_\mathrm{cat}^+}
            + d_{0-}\ket{\psi_\mathrm{cat}^-} \big)
      \notag\\&\qquad+
          \ket{b_1} \otimes \big(d_{1+} \ket{\psi_\mathrm{cat}^+}
            + d_{1-}\ket{\psi_\mathrm{cat}^-} \big)
        \Big)
    \notag\\
      &= \frac{1}{\sqrt{2}}\Big(
          \e{i\Theta_0}\ket{b_0}
              \otimes \big(\cos{\varphi} \ket{\psi_\mathrm{cat}^+}
          + \sin{\varphi} \e{i\theta}\ket{\psi_\mathrm{cat}^-} \big)
    \notag\\&\qquad+
          \e{i\Theta_1}\ket{b_1}
              \otimes \big(\sin{\varphi} \ket{\psi_\mathrm{cat}^+}
          - \cos{\varphi} \e{i\theta}\ket{\psi_\mathrm{cat}^-} \big)
        \Big).
    \label{eq:rewriting_general_cat_arb_qubit_comp_entg}
  \end{align}
  Since $\braket{\psi_{\mathrm{0,a^2}}|\psi_{\mathrm{1,a^2}}}=0$, we can
  reexpress the prefactors of the cat states as
  \begin{alignat*}{3}
    d_{0+} &= \e{i\Theta_0}\cos\varphi,
    &&\qquad\quad
    d_{0-} &&=\phantom{-}\e{i\Theta_0}\e{i\theta}\sin\varphi,
    \\
    d_{1+} &= \e{i\Theta_1}\sin\varphi,
    &&\qquad\quad
    d_{1-} &&= -\e{i\Theta_1}\e{i\theta}\cos\varphi,
  \end{alignat*}
  with $\varphi,\theta,\Theta_i \in \mathbb{R}$.
  Using these relations, we rewrite \eqref{eq:rewriting_general_cat_arb_qubit_comp_entg} as
  \begin{align}
    \ket{\Psi_{\mathrm{ent}}}
      &= \frac{1}{\sqrt{2}}\Big(
          \e{i\Theta_0}\ket{b_0}
              \otimes \big(\cos{\varphi} \ket{\psi_\mathrm{cat}^+}
          + \sin{\varphi} \e{i\theta}\ket{\psi_\mathrm{cat}^-} \big)
    \notag\\&\qquad+
          \e{i\Theta_1}\ket{b_1}
              \otimes \big(\sin{\varphi} \ket{\psi_\mathrm{cat}^+}
          - \cos{\varphi} \e{i\theta}\ket{\psi_\mathrm{cat}^-} \big)
        \Big)
    \notag\\
      &= \frac{1}{\sqrt{2}}\Big(
          \big(
            \e{i\Theta_0}\cos{\varphi}\ket{b_0}
            + \e{i\Theta_1}\sin{\varphi}\ket{b_1}
          \big)\otimes \ket{\psi_\mathrm{cat}^+}
    \notag\\&\qquad+
          \e{i\theta}\big(
            \e{i\Theta_0}\sin{\varphi}\ket{b_0}
            - \e{i\Theta_1}\cos{\varphi}\ket{b_1}
          \big)\otimes \ket{\psi_\mathrm{cat}^-}
    \notag\\
      &\equiv \frac{1}{\sqrt{2}}\Big(\!
            \ket{b_+} \otimes \ket{\psi_\mathrm{cat}^+}
            + \ket{b_-} \otimes \ket{\psi_\mathrm{cat}^-}
        \!\Big).
  \end{align}
  Thus, the state obtained by optimizing the functional in
  \cref{eq:functional_completely_entangled_state} is a entangled cat state as
  defined in \cref{eq:ent_cat_state}.

%###############################################################################
%###############################################################################

\bibliography{references}

\end{document}